# The Impact of the General Data Protection Regulation (GDPR) on Online Tracking

*Klaus M. Miller[1], Karlo Lukic, Bernd Skiera*

Klaus M. Miller, Assistant Professor of Marketing, HEC Paris, Rue de la Libération 1, 78350 Jouy-en-Josas, France, Hi!PARIS Chairholder, Center in Data Analytics & AI for Science, Technology, Business & Society, Phone +33-1-39-67-70-88, millerk@hec.fr.

Karlo Lukic, Post-Doctoral Researcher, Department of Marketing, Faculty of Economics and Business, Goethe University Frankfurt, Theodor-W.-Adorno-Platz 4, 60323 Frankfurt, Germany, Phone +49-17643234998, lukic@wiwi.uni-frankfurt.de.

Bernd Skiera, Professor of Electronic Commerce, Department of Marketing, Faculty of Economics and Business, Goethe University Frankfurt, Theodor-W.-Adorno-Platz 4, 60323 Frankfurt, Germany, Phone +49-69-798-34649, skiera@wiwi.uni-frankfurt.de); also Professorial Research Fellow at Deakin Business School, 221 Burwood Highway, Burwood, VIC 3125, Australia (bernd.skiera@deakin.edu.au).

Acknowledgments:

The authors thank the WhoTracks.me team for providing the data for this research project; participants of the 43rd Annual INFORMS Society for Marketing Science (ISMS) Conference 2021, Munich Summer Institute (MSI) 2021, 19th ZEW Conference on the Economics of Information and Communication Technologies 2021, 17th Symposium on Statistical Challenges in Electronic Commerce Research (SCECR) 2021, Interactive Marketing Research Conference (IMRC) 2021, NBER Economics of Privacy Conference, Spring 2022, European Marketing Academy (EMAC) Conference 2022, 44th Annual INFORMS Society for Marketing Science (ISMS) Conference 2022, 20th ZEW Conference on the Economics of Information and Communication Technologies 2022, VfS Annual Conference 2022, and 2nd DPSN International Data Protection Day 2023; and seminar participants at Goethe University Frankfurt for helpful comments.

Funding:

This project has received funding from the European Research Council (ERC) under the European Union's Horizon 2020 research and innovation program (grant agreement No. 833714). Miller gratefully acknowledges support from the Hi! PARIS Center on Data Analytics and Artificial Intelligence for Science, Business, and Society.

Declarations of interest: none.

---

[1] Corresponding author.

# The Impact of the General Data Protection Regulation (GDPR) on Online Tracking

Abstract

This study explores the impact of the General Data Protection Regulation (GDPR), introduced on May 25th, 2018, on online trackers—vital elements in the online advertising ecosystem. Using a difference-in-differences approach with a balanced panel of 294 publishers, we compare publishers subject to the GDPR with those unaffected (the control group). Drawing on data from WhoTracks.me, which spans 32 months from May 2017 to December 2019, we analyze how the number of trackers used by publishers changed before and after the GDPR. The findings reveal that although online tracking increased for both groups, the rise was less significant for EU-based publishers subject to the GDPR. Specifically, the GDPR reduced about four trackers per publisher, equating to a 14.79% decrease compared to the control group. The GDPR was particularly effective in curbing privacy-invasive trackers that collect and share personal data, thereby strengthening user privacy. However, it had a limited impact on advertising trackers and only slightly reduced the presence of analytics trackers.

# 1.    Introduction

The online advertising market is a cornerstone of the modern digital economy, enabling publishers to monetize their content by delivering targeted advertisements to users. Central to this market are online trackers (hereafter, "trackers")—pieces of software that bundle a specific purpose with tracking functionality.

Tracker providers create trackers, which users, publishers, or advertisers use for specific purposes. For instance, publishers embed trackers into their websites to monitor user behavior, personalize content, enable social media sharing, facilitate commenting features, or deliver targeted advertisements. However, trackers often collect and share user data across multiple publishers and advertisers.

Thus, trackers enable publishers to enhance their content and attract more users, which they can then monetize through targeted advertising. Advertisers use trackers to reach desired audiences more effectively, while tracker providers monetize collected data by offering enhanced services or selling it to third parties. However, trackers also raise substantial privacy concerns because they process personal data, often without the user's explicit knowledge or consent (e.g., Kannan and Li 2017; Beke et al. 2018; Bleier et al. 2020; Lobschat et al. 2021; Wieringa et al. 2021; Verhoef et al. 2022; Beke et al. 2022; Eggers et al. 2023).

Consequently, regulators have stepped in to address users' privacy concerns. The European Union's (EU) General Data Protection Regulation (GDPR), enacted in May 2018, aims to protect user privacy by granting individuals more control over their personal data (European Commission 2016). The GDPR imposes strict data collection and processing rules, significantly impacting how trackers operate within the online advertising market. For example, publishers must process personal data for a specific purpose, such as advertising, with explicit user consent and by collecting no more data than necessary.

Existing literature has extensively explored user privacy concerns (e.g., Eggers et al. 2023; Beke et al. 2022; Wieringa et al. 2021), the functioning and evolution of the market for online trackers (e.g., Mayer and Mitchell 2012; Lerner et al. 2016; Karaj et al. 2018a), and the initial impacts of privacy regulations like the GDPR on online advertising and tracker usage (e.g., Goldfarb and Tucker 2011; Peukert et al. 2022; Johnson et al. 2023). These studies have highlighted the tension between the economic benefits of online tracking and the privacy risks it poses to users. However, there remains a gap in understanding the effects of the GDPR on publishers' use of trackers and how this usage impacts key market actors.

Specifically, prior studies have not fully explored how the regulation affects different categories of trackers aligned with the GDPR's objectives. Some research has examined overall reductions in tracker usage in the short-term (up to three months) post-GDPR and demonstrated a rebound thereafter (e.g., Johnson et al. 2023; Peukert et al. 2022; Lefrere et al. 2024). Still, it is unclear whether the GDPR has reduced publishers' use of trackers in the long term (after six months), especially those that pose higher privacy risks because they collect and share personal data. Such an analysis would allow us to determine whether the GDPR has achieved its intended consequence of reducing highly invasive tracking. Additionally, researchers have not comprehensively assessed how the GDPR affects the ability of all involved actors—users, publishers, advertisers, and tracker providers—to achieve their objectives. Therefore, we need a detailed analysis to understand whether the GDPR achieves its intended consequences and what unintended consequences may have arisen.

This paper examines the impact of the GDPR on online tracking. By categorizing online trackers in a manner that aligns with the GDPR's objectives, we can assess both the intended and unintended consequences of the regulation. By analyzing the period before and after the GDPR's enactment, we determine whether the regulation has effectively reduced the use of trackers that pose higher privacy risks to understand the broader implications for users, publishers, advertisers, tracker providers, and regulators.

Our study makes three main contributions.

First, we introduce a categorization of online trackers that reflects the GDPR's aims, allowing us to analyze the regulation's impact nuancedly. This categorization focuses on dimensions such as purpose and necessity, tracking functionality, type of publisher, and size of tracker provider, which are critical for understanding how the GDPR affects different trackers.

Second, we conceptually describe the market for trackers, detailing the roles of users, publishers, advertisers, and tracker providers, illustrating the conflicting aspects of trackers — providing value for the various actors while raising privacy concerns for users.

Third, we empirically analyze the development of this market before and after the GDPR, using balanced panel data from WhoTracks.me covering 294 publishers over 32 months[1]. We also assess the GDPR's impact on the use of trackers through a difference-in-differences (DiD) analysis, comparing EU publishers subject to the GDPR with non-EU publishers as a control group. By exploring the heterogeneity of the GDPR's effects across different tracker categories, we derive both intended and unintended consequences of the regulation on online trackers.

Understanding these consequences is crucial, as they can inform various actors about the effectiveness of the GDPR and guide future policy decisions. For instance, identifying unintended consequences such as increased market concentration or disproportionate impacts on certain publishers can help regulators refine the regulation to achieve its objectives without causing adverse effects. Additionally, our findings can assist publishers in navigating the balance between compliance and monetization strategies, help advertisers adjust their targeting

---

[1] We extend our analysis in the Web Appendix and show robustness using unbalanced panel data with 29,735 publishers.

approaches, and encourage tracker providers to innovate in response to new regulatory requirements.

The remainder of the paper is structured as follows. Section 2 describes the market for online trackers and the impact of GDPR, including our categorization of online trackers aligned with the GDPR's objectives. Section 3 reviews related literature and situates our work within existing research. Section 4 explains the data and methodology used in our empirical analysis. Section 5 presents the results, highlighting the intended and unintended consequences of the GDPR. Section 6 summarizes the findings and discusses their implications. Finally, Section 7 discusses our study's limitations and suggests future research directions.

## 2. Description of the Market for Online Trackers and the Impact of GDPR

### 2.1. Definition, Purpose, and Tracking Functionality of Online Trackers

#### 2.1.1. Definition of Online Trackers

A tracker is software that combines a specific purpose with tracking functionality. The World Wide Web Consortium (W3C Working Group 2019) defines "tracking" as the "[…] collection of data regarding a particular user's activity across multiple distinct contexts, and the retention, use, or sharing of data derived from that activity outside the context in which it occurred." The tracker provider (e.g., Google) typically develops a tracker, while an installing actor—such as a user, publisher, or advertiser—uses it to serve a particular purpose. For example, a user might install the Google Translate browser extension in his or her browser to view translations instantly as they browse the web. This extension allows the user to access convenient translation services; however, it also includes tracking functionality that enables the tracker provider and sometimes also the publisher or advertiser to collect, retain, or share data on how the user interacts with the extension, sometimes even without the user's explicit awareness.

Similarly, a publisher might install Google AdSense on its website to display targeted ads to users via the Google ad network. Because the publisher and its advertisers want to know which users clicked on ads, ad serving includes tracking functionality that enables such ad measurement on the publisher's website.

### 2.1.2. Purpose of Online Trackers

The purpose of a tracker refers to the aim it serves for the installing actor. For publishers, trackers may serve various purposes, including analytics, advertising, social media integration, and enhancing user experience. For instance, an analytics tracker collects data on user behavior to help publishers optimize website performance and content. Advertising trackers assist publishers and advertisers in displaying targeted ads based on user interests and browsing behavior. Social media trackers enable users to share content across their social media networks, thereby extending the reach of a publisher's content. Understanding the purpose of trackers helps recognize the value they provide to publishers, advertisers, and even users by fulfilling specific needs within the online advertising market.

### 2.1.3. Tracking Functionality of Online Trackers

The tracking functionality of a tracker refers to the specific activities it performs to fulfill its purpose. These activities involve collecting, retaining, using, or sharing data—particularly user data—often across multiple contexts.

For example, an analytics tracker may collect user data, such as browsing history and interaction patterns, to provide insights into how users engage with a publisher's website. Advertising trackers might monitor user behavior across different sites to deliver targeted ads. The tracking functionality defines what the tracker does and how it achieves its intended purpose, often extending beyond the immediate publisher's website to contribute to comprehensive user profiles.

### 2.1.4. Bundling of Purpose and Tracking Functionality

Trackers often bundle their purpose with tracking functionality. While the purpose delivers value by fulfilling specific aims—such as providing analytics, enabling advertising, or enhancing user experience—the tracking functionality involves data collection and processing that may raise privacy concerns for users.

On one hand, the purpose of trackers can bring significant value to publishers, advertisers, tracker providers, and even users. Publishers use trackers to gain valuable insights into user behavior, optimize content, and generate advertising revenue. Advertisers benefit by effectively reaching their target audiences and improving ad relevance and campaign performance. Tracker providers enhance their offerings and maintain a competitive edge through data collected across multiple publishers. Users might enjoy personalized content, relevant advertisements, or features like social media sharing.

On the other hand, the tracking functionality, particularly its ability to collect personal data, introduces significant privacy concerns for users (Kosinski et al. 2013; Wieringa et al. 2021; Beke et al. 2022; Eggers et al. 2023). Detailed tracking can reveal extensive information about a user's identity, interests, and behavior. This revelation can lead to privacy risks such as unwanted surveillance, data breaches, or misuse of personal information. Users may be unaware of the extent of data collection and how firms use or share their information, leading to a lack of transparency and control over their personal data.

These conflicting aspects highlight that bundling of purpose and tracking functionality brings value to the online advertising market but simultaneously raises privacy concerns for users. Addressing this conflict is essential for creating a balanced and privacy-conscious online environment.

## 2.2. *Actors Involved in the Market for Online Trackers*

Trackers sit at the center among four main actors of the online advertising market: users, publishers, advertisers, and tracker providers. Figure 1 illustrates how these actors interact within the online advertising market.

Figure 1: Main Actors Involved in Online Tracking

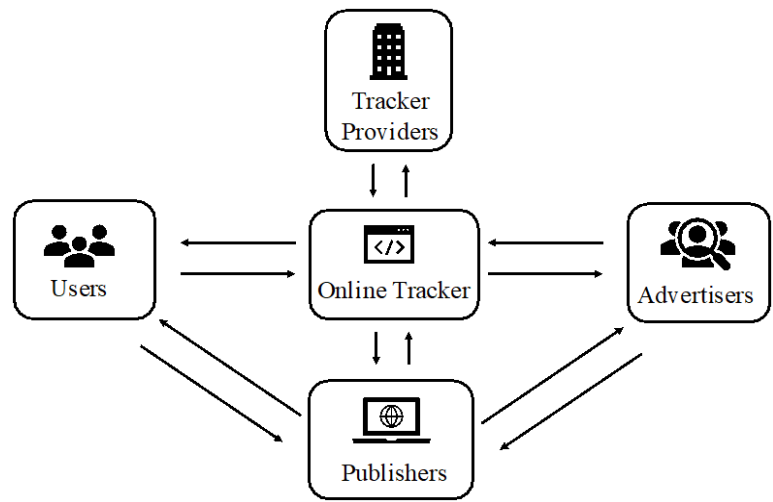

### 2.2.1. Users

Users typically seek access to quality content from publishers, preferably for free, along with a good user experience. Trackers facilitate these goals by enabling publishers to offer free content supported by personalized, targeted ads through advertising trackers. They also enhance the user experience by enabling interactive features, such as social media sharing buttons and comments sections, improving website functionality and performance via content delivery networks, and managing user consent to respect privacy preferences through consent management trackers. However, users may be unaware of the extent to which trackers collect and process their personal data, leading to privacy concerns.

### 2.2.2. Publishers

Publishers rely on trackers to monetize their content and attract and retain users. They embed trackers into their websites for various purposes, including advertising, analytics, social media integration, and enhancing website performance. Advertising trackers help publishers generate revenue by displaying targeted ads. Analytics trackers provide insights into user behavior, aiding in content optimization and website performance improvements. Social media and comment trackers facilitate content sharing and user engagement. Hosting and performance trackers ensure efficient page performance. Additionally, consent management trackers help publishers comply with regulations like the GDPR by respecting user privacy preferences. Publishers must balance their business objectives with the need to protect user privacy and comply with legal requirements.

### 2.2.3. Advertisers

Advertisers aim to reach their target audience and deliver the "right ads to the right users" at the lowest possible cost. Advertising trackers are critical in connecting publishers with advertisers, often through mechanisms like real-time bidding (RTB). These trackers collect user data across publishers' websites and apps for audience profiling, measure user interaction with ads to evaluate campaign effectiveness and the return on investment and employ retargeting and cross-device tracking to increase ad exposure and conversions. Advertisers benefit from the detailed insights trackers provide but must also navigate privacy regulations that may limit data collection and usage.

### 2.2.4. Tracker Providers

Tracker providers develop and supply trackers to publishers, advertisers, and users. They benefit when many publishers use their trackers, as it allows them to collect extensive user data across multiple publishers, increasing the value of the data collected. Additionally, tracker providers profit when publishers, advertisers, or users pay for the use of the tracker. Tracker providers use this data to improve and personalize their services, act as data brokers by selling detailed user profiles to advertisers or other interested parties, and increase their market share by offering trackers fulfilling various purposes. Figure 2 illustrates how tracker providers collect user data across multiple users and publishers.

Figure 2: Tracker Providers Collecting Data Across Users and Publishers

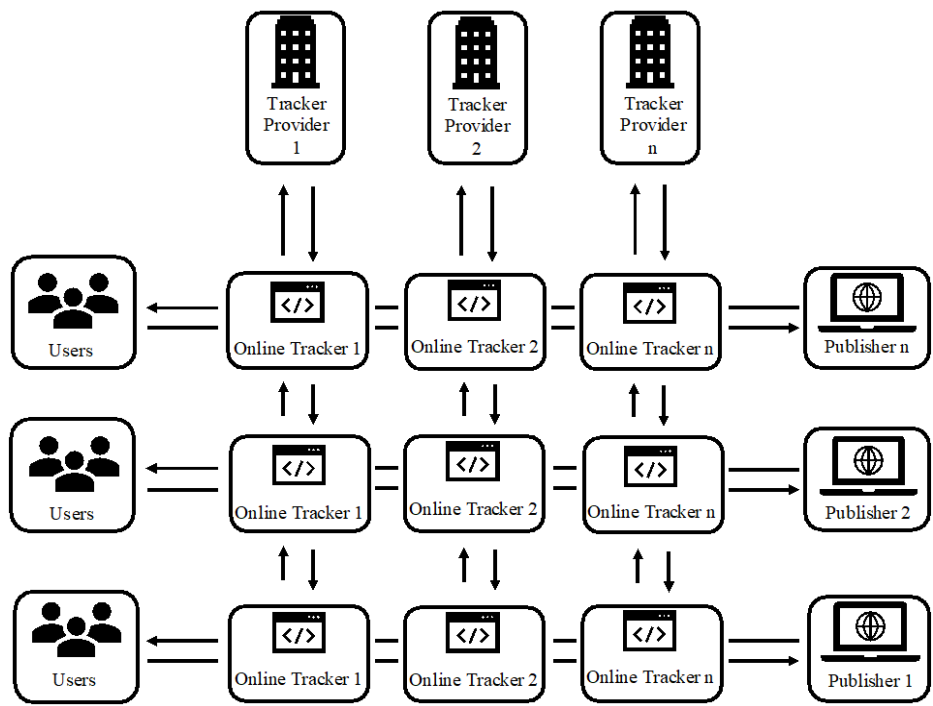

As users navigate across different publishers that employ the same trackers, tracker providers can observe and collect data on these users across publishers. This capability enhances their data collection and market influence. Tracker providers may own multiple trackers, further expanding their presence in the market.

### 2.2.5. *Interplay Among Main Actors in Online Advertising*

Trackers are central to online advertising, facilitating interactions and data flows among users, publishers, advertisers, and tracker providers. Users access content provided by publishers and interact with trackers embedded in websites. Publishers use trackers to monetize content and enhance user experience. Advertisers rely on trackers to collect data for targeted advertising and ad performance measurement. Tracker providers supply the technology that enables data collection and targeted advertising, benefiting all parties but also raising privacy concerns for users. This interconnectedness underscores the complexity of addressing privacy issues without disrupting the value trackers provide to the online advertising market.

### 2.3. *GDPR as a Solution to Increase Online Privacy*

#### 2.3.1. *Aim and Scope of the GDPR*

The GDPR, enacted by EU regulators on May 25th, 2018, is an online privacy law applicable to all EU member states (European Commission 2016). The GDPR aims to increase users' online privacy by strengthening their control over personal data and harmonizing existing national online privacy laws via a single regulation for all EU member states (e.g., Skiera et al. 2022). It achieves these aims by defining users' rights concerning their personal data—such as accessing, editing, or removing data—and imposing obligations on firms that process such data, including tracker providers and publishers that embed trackers on their websites.

The GDPR pertains to the collection and processing of "personal data," which the European Commission (2016) defines as "any information relating to an identified or identifiable natural person" (Article 4). Since online tracking involves creating unique identifiers, such as IP addresses and cookie identifiers, representing individuals, any information gathered is considered personal data under the GDPR and is subject to its stipulations.

#### 2.3.2. *GDPR's Impact on the Market for Online Trackers*

Rather than banning tracking outright, the GDPR mandates that tracking must meet one of three legal bases: explicit user consent (Article 6(1)(a), Article 7), contract fulfillment (Article 6(1)(b)), or legitimate interest (Article 6(1)(f)). The GDPR encourages publishers to reassess their tracking practices, focusing on implementing consent management trackers. Publishers can legally track user data only (i) if the user has explicitly provided consent to be tracked, (ii) if tracking is necessary for providing the requested service, or (iii) if a legitimate interest exists that justifies tracking.

The GDPR imposes obligations on publishers to conduct Data Protection Impact Assessments (DPIAs), which audit their data processing practices. These assessments require publishers to examine how personal data is collected, used, and shared and to determine whether these practices meet the GDPR standards. As a result, publishers may choose to remove or adjust specific trackers to reduce the risk of non-compliance or rely more heavily on consent management trackers to streamline compliance efforts. GDPR also imposes significant penalties for non-compliance, including fines of up to €20 million or 4% of the firm's global annual revenue, whichever is higher (European Commission 2016). These stringent measures incentivize firms to prioritize data protection and foster a culture of privacy by design.

### 2.3.3. *Regional Applicability and Implications*

Previous EU privacy laws only affected firms based in the EU. In contrast, the GDPR applies to all firms that process EU citizens' personal data. It only treats EU and non-EU firms differently in processing non-EU citizens' personal data; in that case, the GDPR applies to EU firms but does not apply to non-EU firms. Thus, a non-EU publisher that receives traffic and processes data from both EU and non-EU users must comply with the GDPR when processing data of EU users but is not obligated to apply the same standards for non-EU users (European Data Protection Board 2018). This broad applicability has significant implications for global firms operating in the online advertising market.

### 2.4. *Categorization of Online Trackers*

The categorization of online trackers is essential for understanding how different types of trackers contribute to both the value provided in the online advertising market and the privacy concerns they raise. More importantly, it allows us to analyze how the GDPR, as a regulatory intervention, impacts various categories of trackers differently, shedding light on both the intended and unintended consequences of the GDPR. By aligning the categorization of online trackers with the GDPR's objectives and societal desirability, we can better assess whether the regulation achieves its goals.

The GDPR aims to enhance user privacy by limiting unnecessary and intrusive data collection, requiring explicit user consent for non-essential data processing activities, ensuring that data collection is proportional and necessary for the service provided, and encouraging transparency and accountability among data processors (European Commission 2016). From a societal perspective, reducing invasive tracking practices that compromise user privacy without providing substantial benefits to users is desirable. Essential services that users expect should function without exposing them to unnecessary privacy risks. Therefore, our categorization focuses on dimensions that reflect these concerns and regulatory objectives.

We categorize trackers across five dimensions: purpose, necessity, tracking functionality, type of publisher, and size of tracker provider. These dimensions capture critical aspects relevant to publishers and regulators, particularly under the GDPR, and allow for a compelling description of the tracker market and an understanding of how the GDPR has impacted it.

### 2.4.1. Online Trackers by Purpose and Necessity

Categorizing trackers by their purpose helps us understand their function for publishers. Trackers serve various purposes, such as analytics, advertising, social media integration, and consent management. For example, analytics trackers collect data to optimize website performance, while advertising trackers display targeted ads based on user behavior.

We also categorize trackers by necessity, distinguishing between essential and non-essential trackers. Essential trackers support the basic functionality of a website, such as security, core features, or improving page load times. Under the GDPR, these trackers are allowed without user consent under the legal basis of "legitimate interest" because they are crucial for providing the requested service. Non-essential trackers enhance user experience or provide additional insights but are not critical for functionality. Publishers must justify their use under the GDPR, usually through explicit user consent (European Data Protection Board 2019).

This categorization by purpose and necessity allows us to assess whether the GDPR reduces the use of non-essential trackers that pose higher privacy risks, such as advertising or analytics trackers.

From a societal perspective, essential trackers are acceptable since they are needed to provide the services users request. Reducing non-essential trackers minimizes privacy risks and aligns with users' expectations for privacy protection. Regulators like the French data protection authority CNIL (2023) provide guidelines on using trackers by purpose and necessity. By categorizing trackers this way, we can analyze whether the GDPR leads to a decrease in non-essential tracking, thus achieving its intended consequences. Table 1 categorizes online trackers by purpose and necessity (adapted from Karaj et al. 2018a).

Table 1: Categorization of Online Trackers by Purpose and Necessity

| Purpose | Description of Purpose | Examples of Trackers | Defined By | Necessity | Description of Necessity |
|---|---|---|---|---|---|
| Privacy-Friendly Site Analytics | Collects and analyses data related to website usage and performance. | Piwik Pro, eTracker, eStat | CNIL | Essential | Strictly necessary for the basic functionality of the website.<br><br>Exempt from user consent requirement under GDPR. |
| Tag Managers, Error Reports and Performance | Site requests that may be critical to website functionality, such as tag manager, privacy notices, error reports and performance. | Google Tag Manager, Google Recaptcha, Adobe Typekit | WhoTracks.me | | |
| Consent | Cookie consent managers allow websites to track different levels of user activity. | OneTrust, Cookiebot, IAB Consent | WhoTracks.me | | |
| Content Delivery Network (CDN) | Delivers resources for different site utilities and usually for many other customers. | Amazon CDN, CloudFlare, jQuery | WhoTracks.me | | |
| Hosting | Service used by the content provider or site owner. | GitHub Pages, FastPic, Amazon CloudFront | WhoTracks.me | | |
| Advertising | Provides advertising or advertising-related services such as data collection, behavioral analysis, or re-targeting. | DoubleClick, ShareThis, Experian Marketing Services | WhoTracks.me | Non-Essential | Not strictly necessary for the basic functionality of the website.<br><br>Not exempt from user consent requirement under GDPR. |
| Site Analytics | Collects and analyzes data related to website usage and performance. | Google Analytics, Adobe Analytics, Hotjar | WhoTracks.me | | |
| Social Media | Integrates features related to social media sites. | Facebook Social Plugins, Giphy, Twitter | WhoTracks.me | | |
| Comments | Enables comments sections for articles and product reviews. | Disqus, eKomi, Livefyre | WhoTracks.me | | |
| Audio Video Player | Enables websites to publish, distribute, and optimize video and audio content. | YouTube, Twitch, Spotify | WhoTracks.me | | |
| Miscellaneous | This tracker does not fit into other categories. | Autoscout24, Oracle RightNow, Vinted | WhoTracks.me | | |
| Customer Interaction | Includes chat, email messaging, customer support, and other interaction tools. | PayPal, Google Translate, LiveChat | WhoTracks.me | | |
| Unknown | This tracker has either not been labeled yet or does not have enough information to label it. | boudja.com, xen-media.com, statsy.net | WhoTracks.me | | |

## 2.4.2. Online Trackers by Tracking Functionality

We categorize trackers based on whether they collect personal data, share personal data, or both. This categorization aligns closely with the GDPR's emphasis on collecting, processing, and protecting users' personal data. Trackers that collect personal data might gather information such as IP addresses, browsing history, or other identifiers that tracker providers can use to track user behavior across different publishers. Trackers that share personal data can transmit user information to other entities, such as advertising networks or data brokers, for purposes like targeted advertising and user profiling. By focusing on tracking functionality, we can better understand the level of privacy intrusion associated with different trackers and assess their compliance with the GDPR. This assessment allows us to evaluate whether the GDPR effectively reduced high-risk trackers, such as trackers that collect and share personal data, and thus answer whether the GDPR better protects user privacy, an intended consequence of the regulation. From a societal desirability standpoint, reducing personal data collection and sharing minimizes privacy risks and potential misuse.

### 2.4.3. Online Trackers by Type of Publisher

Different types of publishers have distinct business models and goals, influencing their choice and use of trackers. For example, news publishers may rely heavily on advertising revenue and thus use more advertising trackers to monetize their content (Libert and Nielsen 2018). In contrast, non-news publishers like e-commerce sites or blogs might focus more on analytics trackers to optimize user experience and drive sales. Understanding how different publishers use trackers helps analyze the market more effectively and assess the varying impacts of the GDPR across different types of publishers.

The motivation for this categorization lies in the GDPR's uniform application across different types of publishers, but with the recognition that different publishers may have varying capabilities to adapt. From a societal perspective, it is essential to ensure that vital services, like news, remain accessible without compromising user privacy. By examining publisher types, we can identify if the GDPR disproportionately affects specific sectors, potentially leading to unintended consequences such as reduced content availability or increased financial pressure on certain publishers.

### 2.4.4.  Online Trackers by Size of Tracker Provider

We categorize trackers by the size of the tracker provider, focusing on whether the provider has a high or low market share. High market share providers are often associated with well-known firms familiar to users, such as Google and Facebook, offering multiple trackers and dominating the market. These providers typically own multiple trackers, leading to broad influence across the tracker market.

In contrast, low market share providers are often less well-known firms. Examples include providers like [24]7 or Accord Group AdMicro, whose trackers are used by fewer publishers, resulting in a lower market share. These tracker providers may own a single tracker, limiting their influence compared to high market share providers.

The motivation for this categorization stems from the GDPR's aim to promote fair competition and prevent dominant players' misuse of personal data. From a societal desirability perspective, avoiding market concentration that could reduce consumer choice and innovation is crucial. By analyzing the impact on tracker providers of different sizes, we can assess whether the GDPR inadvertently benefits larger firms that are able to absorb compliance costs, leading to unintended consequences like increased market concentration. This analysis is critical for understanding the competitive dynamics within the tracker market post-GDPR.

### 2.4.5.  Summary of Categorization of Online Trackers

By aligning our categorization of online trackers with the GDPR's objectives and societal desirability, we set up our empirical study to directly examine whether the GDPR achieves its intended consequences and what unintended consequences may have emerged. This categorization enables us to measure changes in tracker usage across dimensions relevant to the GDPR's aims, assess the effectiveness of the GDPR in reducing non-essential and high-risk trackers, and identify any unintended market consequences, such as increased market concentration or disproportionate impacts on certain publishers or tracker providers.

This framework allows us to derive both the intended and unintended consequences of the GDPR on online trackers, providing valuable insights for policymakers, regulators, and actors in the online advertising market.

## 3. Related Literature

### 3.1. Literature on User Privacy Concerns

The first stream of literature that informs our work focuses on user privacy concerns. User privacy concerns are shaped by a complex interplay of perceived risks, contextual expectations, and the perceived benefits of data sharing. Eggers et al. (2023) and Beke et al. (2022) highlight the concept of a "privacy calculus" where users balance privacy risks against the advantages they receive, such as personalization or convenience. This calculus often leads to a "privacy paradox," where users express high privacy concerns but still share their data when offered compelling incentives (Beke et al. 2022). Users frequently prioritize the perceived value they gain over privacy, especially in contexts where data collection leads to more personalized experiences or tangible rewards (Bleier et al. 2020). Additionally, Schumacher et al. (2023) demonstrate that national cultural differences moderate trade-offs between privacy concerns and perceived benefits.

Moreover, privacy concerns are further complicated by how users perceive the context of data use. Lobschat et al. (2021) and Martin et al. (2017) discuss "contextual integrity" as a critical concept shaping privacy expectations. Users are comfortable sharing data if its collection and usage align with their understanding of the context. When breaching these expectations, privacy concerns increase significantly. Transparency and control are, therefore, vital in managing privacy concerns. As Wieringa et al. (2021), Gopal et al. (2023), and Kannan and Li (2017) emphasize, giving users straightforward information about how their data is collected, shared, and used, and providing them with control over their privacy settings, can alleviate privacy-related anxieties. Beke et al. (2018) also underline the importance of providing users with mechanisms to control their data, which can mitigate privacy concerns.

The impact of personalized advertising on user privacy has also been a focal point of discussion. Schumann et al. (2014) and Tucker (2012) found that personalized ads often heighten privacy concerns, particularly when users feel unaware or uninformed about firms using their data ad targeting despite the ads' potential value. Ahmadi et al. (2024) further illustrate that extensive targeting in digital marketing may lead to a sense of discomfort and concern among users about data misuse. These findings highlight that firms must improve transparency in personalized advertising and data use. Verhoef et al. (2022) suggest that firms balance personalization benefits with privacy considerations to ensure user trust.

Corporate responsibility plays an essential role in mitigating privacy concerns. Lobschat et al. (2021) discuss Corporate Digital Responsibility (CDR) as a means for companies to address user privacy proactively. Companies that are transparent about their data practices and demonstrate accountability in handling user data can build trust and reduce privacy concerns. Users also express unease with data sharing among third parties, especially when the sharing lacks transparency, as highlighted by Gopal et al. (2018) and Martin et al. (2017). Rocher et al. (2019) further amplify these concerns, showing that advancements in data analytics make re-identification from supposedly anonymized datasets increasingly feasible, exacerbating fears about privacy (see also Sweeney 2000; Dinur and Nissim 2003).

In summary, the key insights from these studies underscore user privacy concerns are linked to how transparent and ethical companies are in their data practices. Users want more control and understanding over their data, especially when targeted advertising and data sharing are prevalent. By improving transparency, providing meaningful user control, and demonstrating corporate responsibility, firms can better navigate users' privacy concerns while balancing the value derived from data-driven marketing practices.

### 3.2.  *Literature Describing the Market for Online Trackers*

The second stream of literature that informs our work focuses on the market for online trackers, exploring main parties involved, the services offered, and the economic and privacy implications of tracking technologies. Mayer and Mitchell (2012) conducted a comprehensive survey of third-party online trackers, highlighting the roles of trackers, publishers, advertisers, and tracker providers in online advertising. They identified business models that include ad networks, ad exchanges, and analytics services like Google Analytics, which enable detailed user profiling, often raising significant privacy concerns. The study emphasized the tension between the economic benefits of tracking, such as funding free content, and the privacy risks posed to users, with large tracker providers such as Google Analytics, DoubleClick, and Facebook being significant contributors.

Lerner et al. (2016) conducted a longitudinal analysis of third-party online trackers, revealing a significant increase in tracking between 1996 and 2016. By 2016, many publishers had used 20 or more trackers, with Google Analytics present in nearly 46% of all measured web traffic and about 70% of the top publishers embedding Google-owned trackers. This study highlighted the growing market concentration of tracking within a small group of dominant tracker providers, notably Google.

Karaj et al. (2018b) analyzed the market for online trackers before the GDPR, analyzing data from 1.5 billion web pages. They found that tracking was pervasive, with an average of eight trackers per publisher and 71% of user traffic containing tracking elements. Google's trackers were present in 82% of web traffic, emphasizing the dominance of a few major players like Google, Facebook, and Amazon. The study noted the continued evolution of tracking technologies and underscored the concentration of market power among these large tracker providers.

### 3.3. Literature on the Impact of Privacy Regulation on the Market for Online Trackers

The third stream of literature that informs our work examines the impact of privacy regulations, particularly the GDPR, on the market for online trackers. Goldfarb and Tucker (2011) found that the e-Privacy Directive led to a 65% reduction in advertising effectiveness, especially for general content publishers and simple ads, by limiting the ability to use behavioral data. So, privacy regulations can reduce the effectiveness of targeted advertising and potentially lead to shifts towards more obtrusive ad formats.

Peukert et al. (2022) studied the effects of the GDPR on tracker usage, observing significant reductions among EU publishers and even non-EU publishers targeting EU users, a phenomenon known as the "Brussels effect." Despite a 7.9% reduction in tracker usage by EU publishers and a 12.5% reduction by non-EU publishers, market concentration increased, with Google benefiting disproportionately. So, privacy regulations may unintentionally reinforce the dominance of large tracker providers.

Similarly, Johnson et al. (2023) found a 15% reduction in tracker usage post-GDPR and a 17% increase in market concentration, indicating that larger tracker providers could better absorb compliance costs, further entrenching their market positions. Godinho De Matos and Adjerid (2022) found that GDPR-compliant consent mechanisms improved user consent rates, leading to a 16.1% increase in targeted marketing effectiveness and a 3.7% rise in sales for publishers, showing how transparent consent processes can have positive economic outcomes.

Wang et al. (2024) reported modest negative effects of GDPR on advertising, such as a 2.1% decrease in click-through rates and a 5.7% drop in revenue per click. However, the study noted that contextual targeting was a viable alternative to behaviorally targeted ads in a privacy-regulated environment. Goldberg et al. (2024) found that the GDPR led to a 12% reduction in page views and a 13% decline in e-commerce revenue, highlighting significant trade-offs between privacy protections and user engagement.

Lefrere et al. (2024) observed a 27.4% reduction in cookie-based tracking among EU publishers but no significant effect on user engagement or content provision compared to US publishers. So, EU publishers have adapted effectively to GDPR requirements without severely impacting user engagement. Lastly, Miller and Skiera (2024) analyzed the potential economic consequences of proposed privacy regulations, noting that limiting cookie lifetime could reduce ad revenue by up to €904 million annually, underscoring the need for careful cost-benefit analyses before implementing further privacy restrictions.

### 3.4. *Contribution of our Study to the Related Literature*

Our study contributes to these three literature streams in several ways:

First, we introduce a categorization of online trackers that aligns with the GDPR's objectives, outlining key actors and the value of online trackers for each. This categorization allows us to analyze how the GDPR impacts different types of trackers differently, deriving both intended and unintended consequences of the regulation.

Second, we empirically analyze the market for online trackers from 2017 to 2019, providing a detailed examination of tracker usage patterns.

Third, we examine how the GDPR impacted the market for online trackers, focusing on its effects on tracker usage and user privacy concerns. Specifically, we explore market heterogeneity by tracker purpose and necessity, tracking functionality, publisher type, and tracker provider size, extending our understanding of the GDPR's impact on different categories.

Regarding the literature on user privacy concerns, we examine publishers' privacy practices—particularly the large-scale information collection outlined by Beke et al. (2018)—by analyzing the online tracker market and the GDPR's impact on giving users increased control over their data. We empirically assess how the GDPR influences tracker usage, especially in reducing non-essential and personal-data-collecting trackers. By identifying both intended reductions in high-risk trackers and unintended market shifts, we provide empirical evidence of the regulation's effectiveness and broader implications, addressing gaps in the literature. Our results support the effectiveness of privacy regulations in mitigating concerns, as noted by Martin et al. (2017) and Gopal et al. (2023). We add depth by examining how different types of publishers respond to regulations and how these responses impact different types of trackers, aligning with the contextual privacy discourse discussed by Beke et al. (2022) and Lobschat et al. (2021). Furthermore, our findings offer empirical support for Corporate Digital Responsibility (CDR), showing how the GDPR drives firms towards responsible data practices.

Regarding the literature describing the market for online trackers, our study highlights how the GDPR has reshaped tracker usage across publisher types, enhancing understanding of market dynamics following regulatory interventions. We extend prior studies by examining data from 2017 to 2019, whereas earlier research focused on older periods. Building on the work of Mayer and Mitchell (2012) and Lerner et al. (2016), we show how regulations like the GDPR disrupt established tracking patterns and influence market dynamics without necessarily increasing market concentration.

Building upon the literature on the impact of privacy regulation on online trackers, we extend prior studies (e.g., Peukert et al., 2022; Johnson et al., 2023; Lefrere et al., 2024) that primarily relied on web crawlers simulating user behavior. In contrast, we utilize data from WhoTracks.me, sourced from real users who reported the trackers they encountered. By categorizing this user-based data, we capture actual user exposure to different trackers, providing a more nuanced assessment of the GDPR's effects on user privacy. This assessment reflects users' privacy concerns, as indicated by the number of trackers on websites.

## 4.    Setup of Empirical Study

### 4.1.    *Description of the Data Sets*

Table 2 provides an overview of the data sets used in our study, highlighting their source, the type of information they contain, the periods they cover, and their purpose in reaching the aim of our study.

WhoTracks.me is our primary data source, offering detailed information on 294 publishers' use of trackers over 32 months from May 2017 to December 2019. (i.e., 12 months pre-GDPR and 20 months post-GDPR; Karaj et al. 2018b). We ended our observation period in December 2019 because the California Consumer Privacy Law (CCPA) was enacted in January 2020, and its effects might have interacted with those of the GDPR, confounding our observations from that month onward. Additionally, we chose to use a balanced panel of publishers for our main analysis to avoid panel attrition. In Web Appendix 9.15, we show the robustness of our results using an unbalanced panel with 29,735 unique publishers.

The WhoTracks.me data allows us to empirically describe online trackers and assess the impact of the GDPR on them. We use various information from WhoTracks.me, including the number of trackers and categorizations by purpose, necessity, and size of tracker provider. Additionally, the data includes information that enables us to designate publishers as EU vs. non-EU based on their top-level domain. In Web Appendices 9.6-9.79.8, we detail the raw data available from WhoTracks.me, how WhoTracks.me collects its data and how we prepared our sample of publishers, respectively.

We augment the WhoTracks.me data with publicly available data from SimilarWeb. This data provides traffic shares from the top five countries (EU and non-EU) for each of 294 publishers as of August 2021, which we use to refine our EU vs. non-EU publisher designations further.

Lastly, we augment the WhoTracks.me data with publicly available data from Evidon, which provides additional information on trackers from their privacy policies. This data enables us to categorize trackers by their tracking functionality, particularly regarding personally identifiable information (PII) data collection and sharing practices. Evidon data includes information on 724 trackers (76%) of the 949 unique trackers from the WhoTracks.me data, offering detailed insights into their PII data collection and sharing practices.

Table 2: Description of the Data Sets

| Data Set | Source | Contained Information | Period | Purpose |
|---|---|---|---|---|
| WhoTracks.me | Public | - Publishers' use of trackers<br>- Trackers (e.g., purpose, tracker provider)<br>- Monthly data for 294 publishers over 32 months<br>- Balanced panel of 9,408 observations (294 publishers * 32 months)<br>- Information about publisher types<br>- For each publisher, top-level domain used to categorize as EU vs. Non-EU, in combination with SimilarWeb data | 05/2017 - 12/2019 | Main data set to empirically describe online trackers and measure impact of GDPR's enactment on online trackers |
| SimilarWeb | Public | - Traffic shares from the top five (EU and non-EU) countries<br>- Information on 294 out of 294 (100%) publishers in the balanced panel | 08/2021 | Augments WhoTracks.me data set to categorize publishers as EU vs. Non-EU based on majority of traffic shares |
| | Proprietary | - Daily-level information on traffic shares for 7,332 publishers<br>- Traffic shares for US users and specific EU countries<br>- Information on 200 out of 294 (68%) publishers in the balanced panel | 01/2018 - 12/2019 | Augments public SimilarWeb data set to check the consistency of publisher's website traffic distribution over time |
| Evidon | Public | - Information on trackers from their privacy policies<br>- 724 (76%) matched trackers of 949 unique trackers from WhoTracks.me<br>- 546 (75%) of 724 disclose data collection and sharing practices<br>- 35 (4%) disclose only data sharing, 0 disclose only data collection, 143 (15%) disclose neither practices<br>- 225 (24%) trackers do not match | 03/2021 | Augments WhoTracks.me data set to categorize trackers based on tracking functionality from their disclosed data collection and sharing practices |

## 4.2. Description of the Number of Trackers

For each publisher in our WhoTracks.me data, we measure the number of trackers in a particular month and use it as our dependent variable in the subsequent analysis. Tracing the number of trackers allows us to document the publisher's privacy practices—particularly the large-scale information collection outlined by Beke et al. (2018)—and ultimately serves as a measure of a user's exposure to privacy risk.

To obtain a reliable measurement of the number of trackers initiated by our focal publishers, we adjusted the raw number of trackers provided by WhoTracks.me, as follows.

First, WhoTracks.me counts specific browser extensions (e.g., Kaspersky Labs, Adguard) that users voluntarily install as "trackers". WhoTracks.me identifies these trackers as "extensions". As users install these trackers rather than the publishers themselves, we excluded them from each publisher's overall tracker count.

Second, we counted only third-party trackers because the European Data Protection Authorities (DPAs) consider them a greater privacy risk than first-party trackers (Article 29 Data Protection Working Party 2012). We defined these as trackers whose tracker providers differed from the publisher. Our definition of third-party trackers closely follows that of the European DPAs (Article 29 Data Protection Working Party 2012).

We acknowledge that the set of trackers on a publisher's website that WhoTracks.me reports may not be fully comprehensive. Specifically, WhoTracks.me does not capture certain types of trackers, so we cannot include them in our dataset. As Ghostery (2017) noted, these trackers are typically found on fewer than ten publishers or do not rely on cookies or fingerprinting technologies to track user identifiers. Consequently, we do not expect these trackers' omissions to significantly impact our results' reliability.

To explore the effects of the GDPR on trackers, we also count, for each publisher and each month, the numbers of trackers corresponding to the tracker categorizations as elaborated in Section 2.4.

### *4.3.    Construction of the Treatment and Control Groups*

We broadly define our treatment group as publishers subject to the GDPR; as discussed above, these publishers corresponded to EU firms. We broadly define our control group as publishers not subject to the GDPR—i.e., non-EU firms.

Determining whether a publisher represents an "EU firm" or a "non-EU firm" is challenging. The GDPR defines an "EU firm" as any firm established within the EU, including firms based in the EU or that process EU citizens' personal data—regardless of where the firm is based. In contrast, the GDPR defines a "non-EU firm" as any firm that is not established within the EU and does not process the personal data of EU citizens (European Commission 2016; European Data Protection Board 2018).

We suggest several proxies to identify whether a publisher is an "EU" or a "non-EU firm": the publisher's (1) target audience (i.e., the main set of users it caters to), (2) (country-code) top-level domain (TLD), (3) cookie banner display, (4) server location, and others. Our main analysis combines two proxies: the publisher's target audience and its TLD.

Specifically, we identify a publisher as an "EU publisher" if it fulfills one of the following criteria: (1) the publisher's TLD includes an EU country code (e.g., .de); *or* (2) the publisher receives more traffic from EU users than from non-EU users in at least one month (e.g., August 2021), according to our SimilarWeb data. Correspondingly, a publisher is designated as a "non-EU publisher" if it fulfills *both* of the following criteria: (1) the publisher uses a non-EU TLD (e.g., .com), and (2) the publisher receives more traffic from non-EU users than from EU users.

Our sample corresponds to 294 EU and non-EU publishers over 32 months (N = 9,408 observations). Table 3 shows how we assign the different publishers and their corresponding monthly observations to the treatment and control groups. In total, 22.79% (N observations = 2,144) of all observations belong to the treatment group, and 77.21% (N observations = 7,264) belong to the control group.

Table 3: Distribution of Observations (Monthly Publishers) Across Publisher Designation

| Publisher Designation | Number and Percentage of Observations |
|---|---|
| EU publisher[1] | 2,144 (22.79%) |
| Non-EU publisher[2] | 7,264 (77.21%) |
| ∑ | 9,408 (100.00%) |

[1]A publisher is designated as an "EU publisher" if (1) the publisher uses an EU top-level domain (e.g., .de) or (2) the publisher receives more traffic from EU users than non-EU users. [2]A publisher is designated as a "non-EU publisher" if (1) the publisher uses a non-EU top-level domain (e.g., .com) and (2) the publisher receives more traffic from non-EU users than EU users.

Notes: The cells in this table show the number and percentage of observations in our sample corresponding to each case. The cell belonging to the control group—where GDPR does not apply—is colored gray, and the cell belonging to the treatment group—where GDPR applies—is not colored. In total, 23% (N observations = 2,144) of all observations (N observations = 9,408) belong to the treatment group and 77% (N observations = 7,264) to the control group.

### 4.4. Identification Assumptions for the Difference-in-Differences Analysis

We use a DiD analysis to estimate the GDPR's effect on trackers. DiD is suitable because it allows for comparing changes in tracker usage over time between EU publishers (affected by GDPR) and non-EU publishers (unaffected), effectively isolating the GDPR's causal impact. DiD analysis relies on the following two assumptions, the violation of which could bias our results: (1) stable unit treatment value assumption (SUTVA) and (2) parallel trends assumption (Huntington-Klein 2022). In what follows, and following Goldfarb et al. (2022), we discuss these two assumptions in the context of our study.

SUTVA comprises two parts. The first part states that there is no hidden variation of treatment, i.e., that all treated publishers receive the same level of treatment. This assumption is fulfilled, as the treatment we observe is the GDPR's enactment—which imposes consistent requirements across all firms subject to the GDPR.

The second part of SUTVA states that no spillover effects exist between the treatment and control groups. We stringently define our control group to avoid such spillover (see description above and Table 3). We acknowledge, however, that, despite not being required to do so, some publishers in our control group might have altered their tracking practices following the GDPR to avoid inadvertently incurring the GDPR fines, thus "contaminating" our control group. This concern is present in most, if not all, studies on the impact of the GDPR (e.g., Johnson 2023), and we address it in Section 5.3.

The parallel trends assumption states that in the absence of the GDPR, the outcome difference between the treatment and control group would have remained the same after the GDPR as before the GDPR. DiD studies commonly evaluate this assumption by providing visual evidence of group outcomes in a period before the intervention.

## 5. Results of Empirical Study

In what follows, we empirically describe online trackers and determine the impact of the GDPR on them.

### 5.1. Distribution of Online Trackers

#### 5.1.1. Distribution of Number of Online Trackers Across Publishers

Figure 3 shows the average number of trackers per publisher, averaged across all months in our observation period. It shows that there are, on average, 16.689 trackers per publisher (SD = 13.496). Their distribution is unimodal and right-skewed. Thus, most publishers typically use between 1 and 10 trackers; only a few use many. The largest number of trackers for a publisher in a particular month is 111 trackers. Because the data only includes publishers with at least 1 tracker, the minimum number of trackers is 1.

#### 5.1.2. Distribution of Number of Trackers By Categorizations of Online Trackers

Table 4 provides the distribution of the average number of trackers per publisher across tracker categorizations, providing a baseline understanding of the market for trackers before examining the impact of the GDPR on it.

By necessity, publishers tend to use three times more non-essential trackers than essential ones, with an average of about 12 non-essential trackers compared to around four essential trackers per publisher.

Figure 3: Distribution of the Average Number of Trackers per Publisher

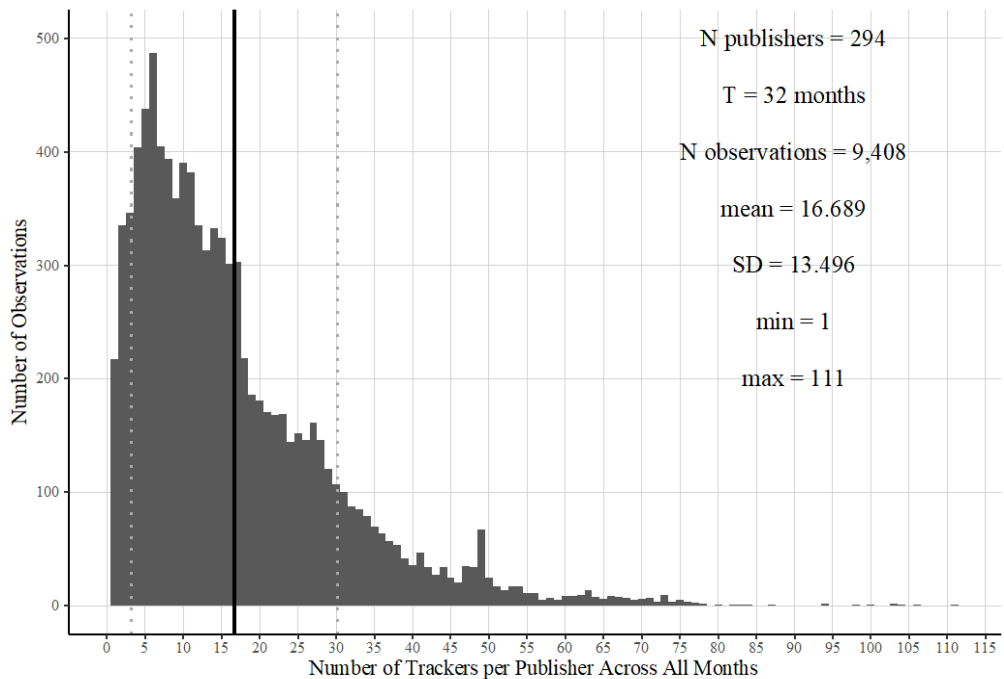

Notes: Multiplying the number of publishers (N publishers = 294) and the number of months (T = 32 months) yields the number of observations (N observations = 9,408). The black vertical line indicates the mean number of trackers per publisher, while the gray lines represent ± one standard deviation from the mean.

By purpose, the distribution indicates that publishers use specific types of essential and non-essential trackers in varying amounts. Publishers commonly use essential trackers such as tag managers and content delivery network (CDN) trackers, with an average of 0.774 tag manager trackers and 2.931 CDN trackers per publisher. For non-essential trackers, publishers mostly use trackers for advertising and analytics purposes, with averages of 7.257 and 2.864 trackers per publisher, respectively. Notably, publishers rarely use privacy-friendly analytics trackers, averaging 0.032 per publisher, suggesting publishers' limited use of these trackers across our entire observation period. Similarly, publishers' use of consent trackers is minimal (0.032 trackers) across the observation period.

Table 4: Distribution of the Average Number of Trackers per Publisher By Categorizations of Trackers

| | Number of Trackers per Publisher Across All Months | | | |
|---|---|---|---|---|
| Categorization of Trackers by Purpose and Necessity | mean | SD | min | max |
| *Essential:* | *4.432* | *3.232* | *0* | *20* |
| Privacy-Friendly Analytics | 0.032 | 0.215 | 0 | 3 |
| Tag Managers, Error Reports and Performance | 0.774 | 0.886 | 0 | 5 |
| Consent | 0.139 | 0.438 | 0 | 4 |
| Content Delivery Network (CDN) | 2.931 | 2.127 | 0 | 12 |
| Hosting | 0.556 | 0.759 | 0 | 4 |
| *Non-Essential:* | *12.258* | *11.125* | *0* | *92* |
| Advertising | 7.257 | 8.103 | 0 | 76 |
| Analytics | 2.864 | 2.386 | 0 | 18 |
| Social Media | 0.659 | 0.937 | 0 | 8 |
| Comments | 0.064 | 0.248 | 0 | 2 |
| Audio Video Player | 0.408 | 0.731 | 0 | 5 |
| Miscellaneous | 0.452 | 0.828 | 0 | 6 |
| Customer Interaction | 0.404 | 0.785 | 0 | 6 |
| Unknown | 0.181 | 0.502 | 0 | 5 |
| Categorization of Trackers by Tracking Functionality | | | | |
| Not Collecting PII | 1.872 | 2.641 | 0 | 25 |
| Collecting PII | 3.258 | 3.553 | 0 | 31 |
| Collecting and Sharing PII | 8.424 | 6.764 | 0 | 46 |
| Unknown (Undisclosed or No Match) | 4.227 | 3.519 | 0 | 32 |
| Categorization of Trackers by Type of Publisher | | | | |
| *News Publishers:* | *28.902* | *19.044* | *1* | *111* |
| News & Portals | 28.902 | 19.044 | 1 | 111 |
| *Non-News Publishers:* | *15.353* | *12.013* | *1* | *103* |
| E-Commerce | 24.971 | 13.166 | 1 | 71 |
| Recreation | 18.781 | 10.405 | 1 | 51 |
| Business | 18.682 | 13.991 | 1 | 77 |
| Entertainment | 16.947 | 12.694 | 1 | 103 |
| Reference | 13.803 | 11.282 | 1 | 78 |
| Adult | 9.545 | 5.075 | 1 | 33 |
| Government | 7.281 | 3.429 | 2 | 11 |
| Categorization of Trackers by Size of Tracker Provider | | | | |
| Trackers of Providers with High Market Share | 8.351 | 5.708 | 0 | 30 |
| Trackers of Providers with Low Market Share | 8.338 | 9.087 | 0 | 83 |

Notes: This table displays descriptive statistics for the number of trackers per publisher across all months and types of tracker categorizations. Italicized labels represent grouped variables, where category descriptives (e.g., "Essential:") are followed by descriptives for subcategories within that group (e.g., "Privacy-Friendly Analytics"). Multiplying the number of publishers (N publishers = 294) and the number of months (T = 32 months) yields the number of observations (N observations = 9,408).

By tracking functionality, the distribution reveals that publishers commonly use trackers that collect and share PII, with an average of 8.424 trackers per publisher. Publishers use trackers that collect but do not share PII less frequently, averaging 3.258 per publisher. In contrast, they use trackers that neither collect nor share PII, least commonly, with an average of 1.872 trackers per publisher.

By type of publisher, the distribution shows that news publishers use almost two times more trackers than non-news publishers, averaging about 29 trackers compared to 15 trackers for non-news publishers. Notably, within the non-news publisher industry, e-commerce publishers stand out with an average of 24.971 trackers per publisher, while government publishers tend to use the fewest, averaging just 7.281 trackers. The distribution reveals that even entertainment publishers who fall under the non-news publisher industry can employ many trackers, with one entertainment publisher using up to 103 trackers in a particular month.

By size of tracker provider, publishers tend to use a similar number of trackers from both high and low market share tracker providers, with averages of 8.351 and 8.338 trackers, respectively, from each type.

### 5.1.3. *Distribution of Number of Trackers By Categorizations of Online Trackers in the Treatment and Control Groups*

Table 5 presents the distribution of the average number of trackers per publisher across the treatment and control groups, categorized by tracker purpose, necessity, functionality, publisher type, and size of the tracker provider.

The treatment group consistently uses more trackers than the control group across most tracker categorizations, with a 31.32% higher overall average (20.457 vs. 15.577 trackers). By purpose and necessity, the treatment group uses more non-essential trackers (35.65% higher), particularly in advertising (58.97% higher). By tracking functionality, the treatment group uses more trackers that collect and share personally identifiable information (19.24% higher). News publishers in the treatment group also use substantially more trackers (54.49% higher) than those in the control group.

In addition to the differences in tracker use between the treatment and control groups, Table 5 also summarizes publisher characteristics.

Table 5: Distribution of the Average Number of Trackers per Publisher By Categorizations of Trackers in the Treatment and Control Groups

| | Treatment Group | Control Group | Difference (%) | |
|---|---|---|---|---|
| Number of Trackers per Publisher Across All Months | 20.457 | 15.577 | 4.879 | (31.32%) |
| Categorization of Trackers by Purpose and Necessity | | | | |
| *Essential:* | 5.078 | 4.241 | 0.838 | (19.75%) |
| Privacy-Friendly Analytics | 0.124 | 0.005 | 0.119 | (2,380.00%) |
| Tag Managers, Error Reports and Performance | 0.857 | 0.750 | 0.107 | (14.28%) |
| Consent | 0.145 | 0.137 | 0.008 | (6.11%) |
| CDN | 3.280 | 2.828 | 0.452 | (15.98%) |
| Hosting | 0.673 | 0.522 | 0.151 | (28.96%) |
| *Non-Essential:* | 15.378 | 11.337 | 4.042 | (35.65%) |
| Advertising | 10.170 | 6.397 | 3.773 | (58.97%) |
| Analytics | 3.049 | 2.810 | 0.239 | (8.51%) |
| Social Media | 0.505 | 0.704 | -0.199 | (-28.25%) |
| Comments | 0.075 | 0.061 | 0.015 | (23.97%) |
| Audio Video Player | 0.443 | 0.397 | 0.045 | (11.45%) |
| Miscellaneous | 0.545 | 0.425 | 0.120 | (28.19%) |
| Customer Interaction | 0.424 | 0.398 | 0.025 | (6.34%) |
| Unknown | 0.292 | 0.149 | 0.143 | (95.71%) |
| Categorization of Trackers by Tracking Functionality | | | | |
| Not Collecting PII | 3.064 | 1.520 | 1.544 | (101.56%) |
| Collecting PII | 3.779 | 3.104 | 0.676 | (21.77%) |
| Collecting and Sharing PII | 9.623 | 8.070 | 1.553 | (19.24%) |
| Unknown (Undisclosed or No Match) | 5.888 | 3.737 | 2.150 | (57.54%) |
| Categorization of Trackers by Type of Publisher | | | | |
| *News Publishers:* | 34.833 | 22.547 | 12.286 | (54.49%) |
| News & Portals | 34.833 | 22.547 | 12.286 | (54.49%) |
| *Non-News Publishers:* | 16.309 | 15.119 | 1.190 | (7.87%) |
| E-Commerce | 25.554 | 24.461 | 1.093 | (4.47%) |
| Recreation | 18.711 | 18.875 | -0.164 | (-0.87%) |
| Business | 29.823 | 18.134 | 11.689 | (64.46%) |
| Entertainment | 17.545 | 16.887 | 0.658 | (3.90%) |
| Reference | 13.812 | 13.800 | 0.012 | (0.09%) |
| Adult | 12.040 | 8.322 | 3.718 | (44.67%) |
| Categorization of Trackers by Size of Tracker Provider | | | | |
| Trackers of Providers with High Market Share | 9.925 | 7.887 | 2.039 | (25.85%) |
| Trackers of Providers with Low Market Share | 10.531 | 7.691 | 2.840 | (36.93%) |
| Publisher Characteristics | | | | |
| Share of Traffic from EU Users | 48.37% | 10.24% | | (38.12 pp) |
| Share of Traffic from Non-EU Users | 15.63% | 43.28% | | (-27.65 pp) |
| 5 Most Common TLDs | com, co.uk, de, fr, net | com, net, org, ru, tv | | |

Notes: This table shows the average number of trackers for the treatment and control groups across all months and types of tracker categorizations. Italicized labels represent grouped variables, where broad category descriptives (e.g., "Essential:") are followed by descriptives for subcategories within that group (e.g., "Privacy-Friendly Analytics"). The table also shows the average share of traffic from (non-)EU users and the five most common TLDs for treatment and control groups. Percent differences are displayed as percentage points (pp) for shares of traffic from (non-)EU users. The Government publisher has been deliberately omitted from this analysis, given that only a single publisher of this type was present in the control group of our sample.

The treatment group has a higher share of traffic from EU users (48.37%) compared to the control group (10.24%), while the control group has a larger share of traffic from non-EU users (43.28% vs. 15.63%). The five most TLDs also differ between the groups, with the treatment group featuring more EU-specific TLDs like ".co.uk" and ".de". The control group, in contrast, commonly uses more global or non-region-specific TLDs like ".com" and ".ru".

## 5.2. Impact of the GDPR on the Market for Online Trackers

### 5.2.1. Change in Number of Online Trackers Before and After the GDPR

To investigate the effect of the GDPR on trackers, we first show how the average number of trackers differs between the treatment and control groups before (May 2017-April 2018) and after the GDPR's enactment (May 2018-December 2019). We run an independent-samples t-test to test whether the group averages differ significantly in the two periods and present the results in Figure 4.

Figure 4: Comparison of the Average Number of Trackers in the Treatment and Control Groups Before and After the GDPR's Enactment

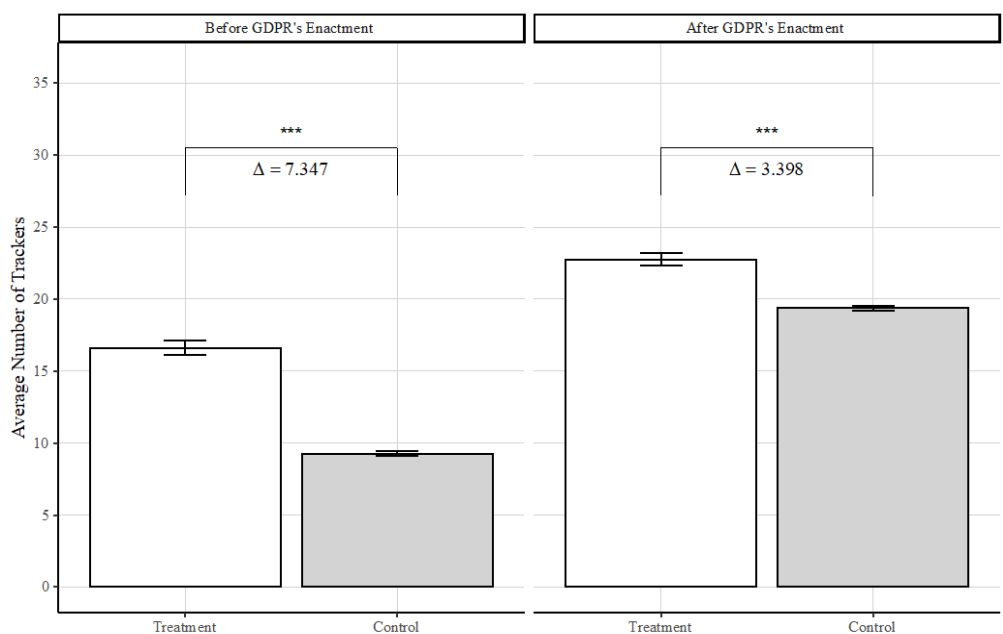

Significance levels: * p < 0.05, ** p < 0.01, *** p < 0.001.
Notes: Error bars = +/− 1 SEs. This figure shows independent t-test comparisons between group averages in periods before (May 2017–April 2018) and after (May 2018–December 2019) the GDPR's enactment using the number of trackers as a dependent variable.

The average number of trackers per publisher was significantly higher in the treatment group than in the control group before the GDPR ($M_{treatment}$ = 16.609 trackers; $M_{control}$ = 9.262 trackers; t(992.7283) = 14.018; p < 0.001). After the GDPR, the number of trackers increased in both groups ($M_{treatment}$ = 22.765; $M_{control}$ = 19.367), with the treatment group still having a significantly higher average number of trackers compared to the control group ($t$(1836.5045) = 6.920; $p < 0.001$).

### 5.2.2.  Change in Number of Online Trackers Over Time

Figure 5 shows how each group's average number of trackers developed over time.

Figure 5: Development of the Average Monthly Number of Trackers in the Treatment and Control Groups

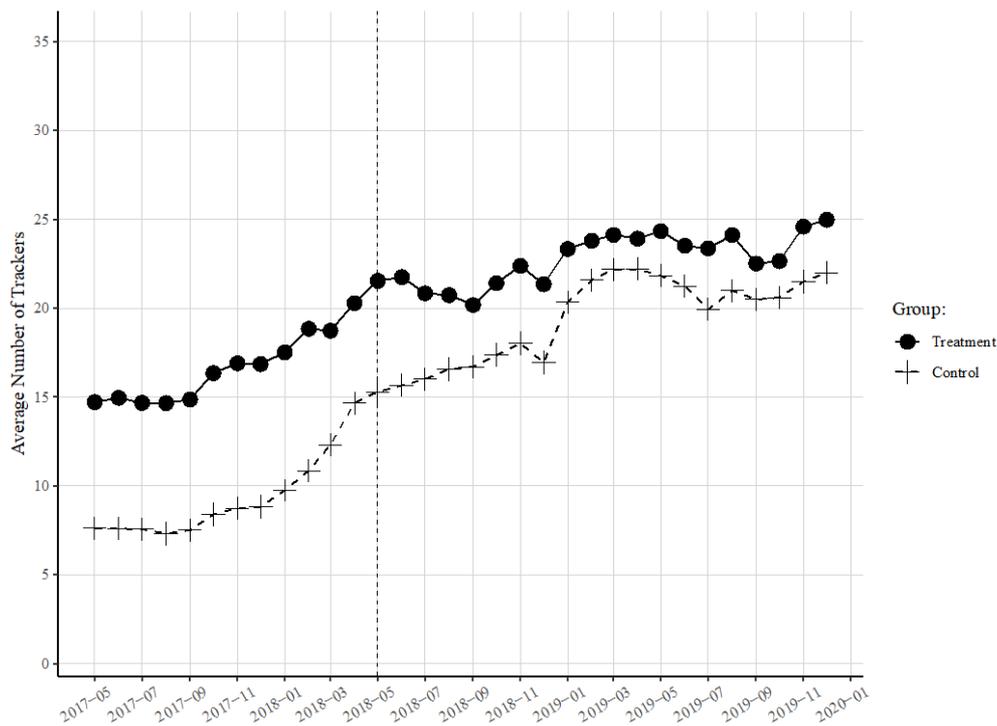

Figure 5 suggests an increasing trend in the number of trackers in the treatment and control groups. The number of trackers in the treatment group was consistently higher than that of the control group throughout the observation period. Around the time of the GDPR's enactment (May 2018), the number of trackers in the treatment group slightly decreased, while the control group increased slightly. However, by the end of 2018, the number of trackers in both groups rises, with the treatment group maintaining a higher level than the control group.

Starting from December 2018, both groups continue to increase the number of trackers, with the treatment group peaking in the last observed month (December 2019) at about 25 trackers, while the control group also increases but remains lower than the treatment group at about 22 trackers.

### 5.2.3. Difference-in-Differences (DiD) Analysis for the Number of Online Trackers

Having obtained a preliminary indication that the GDPR reduced the number of trackers among publishers subject to the regulation (compared with the control group), we examine this effect with a DiD analysis that accounts for unobserved influences.

First, we manually calculate the DiD (average treatment effect on the treated, ATT)–the difference between the average differences in the number of trackers of both groups (Table 6).

Table 6: Cross-Table for the Average (Monthly) Number of Trackers in the Treatment and Control Groups Before and After the GDPR's Enactment

| Group | Before GDPR's Enactment | After GDPR's Enactment | Difference |
|---|---|---|---|
| Treatment | 16.610 | 22.765 | 6.155 (37.06%) |
| Control | 9.262 | 19.366 | 10.104 (109.09%) |
| Difference | 7.347 | 3.398 | -3.949 (14.79%) |

Notes: This table shows the average (monthly) number of trackers for the treatment and control groups in periods before (May 2017-April 2018) and after (May 2018-December 2019) GDPR's enactment and the differences in the average (monthly) number of trackers between groups and periods. We use unrounded values to derive the differences. The values in parentheses represent the percent changes for each group from the period before to the period after the GDPR's enactment. The Difference-in-Differences (DiD) as a percentage is calculated by comparing the observed value in the treatment group after GDPR (22.765) with the expected value if the GDPR had not been enacted. The expected value is calculated by adding the pre-GDPR difference between groups (7.347) to the post-GDPR control group value (19.366), which equals 26.714. The percent decrease is then derived from the ratio of the difference between these two values to the expected value: DiD (%) = $\frac{26.714 - 22.765}{26.714} \times 100 \approx$ 14.79%.

After the GDPR, the average number of trackers increased by 6.155 (37.06%) in the treatment group and 10.104 (109.09%) in the control group. The control group increased the average number of trackers more than the treatment group after the GDPR. The difference between those two numbers captures the effect of the GDPR. It equals -3.949 trackers, suggesting that the GDPR lowered the average number of trackers by about four trackers per publisher.

In other words, these results suggest that if the GDPR had not been enacted, the average publisher in the treatment group would have used about 27 trackers in the post-GDPR period rather than about 23 trackers. This difference corresponds to a 14.79% decrease.

After calculating the DiD, we use an ordinary least squares (OLS) regression to control for other factors that might influence the DiD estimate (e.g., differences between publishers):

$$Y_{i,t} = \alpha + \gamma_i Treatment_i + \delta_t PostGDPR_t + \beta(Treatment_i \times PostGDPR_t) + \epsilon_{i,t} \qquad (1)$$

In Equation (1), our dependent variable for a publisher $i$ at month $t$ is $Y_{i,t}$. $Treatment_i$ is an indicator variable describing whether the publisher $i$ is in the treatment group (i.e., subject to the GDPR; $Treatment_i = 1$) or not ($Treatment_i = 0$). $PostGDPR_t$ indicates the period before the GDPR ($PostGDPR_t = 0$) and the entire period after the GDPR (i.e., after and including May 2018; $PostGDPR_t = 1$). $\gamma_i$ captures the group-specific changes in the outcome variable unrelated to the GDPR (i.e., publisher-fixed effects) and $\delta_t$ captures the time-specific changes in the outcome variable unrelated to the GDPR (i.e., month-fixed effects). $\epsilon_{i,t}$ represents the error term, which includes unobserved factors affecting $Y_{i,t}$. We cluster standard errors at the publisher and month levels to account for autocorrelation within publishers over time and across months. Our coefficient of interest $\beta$ measures the average difference in the number of trackers between both groups over time (i.e., represents the DiD estimate). So, our baseline estimator is a standard two-way fixed effects (TWFE) estimator with the publisher as the observation unit. Table 7 presents the results.

The DiD coefficient ($\beta$ = -3.949, $p < 0.05$, 95% CI [-7.082; -0.816]) is significantly negative in our DiD model presented in column (1). The size of this DiD coefficient is the same as the calculated difference-in-differences in Table 6. These results confirm that the GDPR lowered the number of trackers by about four per publisher (14.79%). We report the estimated publisher and month-fixed effects in Web Appendix 9.12

Table 7: Result of Difference-in-Differences (DiD) Analysis for the Number of Trackers

| Dependent Variable:<br>Model: | Number of Trackers per Publisher and Month<br>(1) |
|---|---|
| Treatment x PostGDPR | -3.949* [-7.082; -0.816] |
| Publisher ID Fixed Effects | ✓ |
| Month ID Fixed Effects | ✓ |
| N Observations | 9,408 |
| $R^2$ | 0.744 |

Significance levels: * $p < 0.05$, ** $p < 0.01$, *** $p < 0.001$.
Two-way standard errors are clustered at the publisher and month levels; 95% confidence intervals are reported in brackets.
Notes: This table shows the difference-in-differences coefficient (Treatment x PostGDPR) from the OLS regression. We assign treatment to each publisher according to the publisher's designation (EU or non-EU). Multiplying the number of publishers (N publishers = 294) and the number of months (T = 32 months) yields the number of observations (N observations = 9,408).

### 5.2.4. Differences in the Impact of the GDPR By Categorizations of Online Trackers

Lastly, we show how the GDPR impacted various tracker categorizations in our sample. We estimate the following model to measure the effect of the GDPR across different tracker categorizations:

$$Y_{i,t}^k = \alpha + \gamma_i Treatment_i + \delta_t PostGDPR_t + \beta(Treatment_i \times PostGDPR_t) + \epsilon_{i,t}^k \qquad (2)$$

In Equation (2), $Y_{i,t}^k$ represents the number of trackers in category $k$ for publisher $i$ in month $t$. The coefficient of interest $\beta$ measures the average difference in the number of trackers in category $k$ between the treatment and control groups over time, representing the DiD coefficient for tracker category $k$. We apply this model separately to three tracker categorizations: by purpose (e.g., advertising, analytics) and necessity (i.e., essential vs. non-essential trackers), by tracking functionality (e.g., trackers that collect and share PII), and by size of tracker provider (i.e., tracker providers with high vs. low market share).

To examine how the GDPR affected broader publisher industries, we use the following model:

$$Y_{i,t}^p = \alpha + \gamma_i Treatment_i + \delta_t PostGDPR_t + \beta(Treatment_i \times PostGDPR_t) + \epsilon_{i,t}^p \qquad (3)$$

In Equation (3), $Y_{i,t}^p$ represents the number of trackers for publisher $i$ in publisher industry $p$ at month $t$. We estimate this model separately for each publisher industry $p$ (i.e., News vs. Non-News). The coefficient of interest $\beta$ captures the impact of the GDPR on the number of trackers within each publisher industry.

Lastly, to examine how the GDPR affected specific types of publishers within these industries, we use:

$$Y_{i,t}^c = \alpha + \gamma_i Treatment_i + \delta_t PostGDPR_t + \beta(Treatment_i \times PostGDPR_t) + \epsilon_{i,t}^c \qquad (4)$$

In Equation (4), $Y_{i,t}^c$ represents the number of trackers for publisher $i$ in publisher type $c$ during month $t$. We estimate this model separately for each specific publisher type $c$ (e.g., News & Portals, E-Commerce, Recreation). The coefficient $\beta$ captures the impact of the GDPR on the number of trackers within each publisher type  (see Web Appendix 9.9 for more details on publisher industries and types).

All three models include $\gamma_i$ publisher-fixed effects and $\delta_t$ month-fixed effects. In all three models, we cluster standard errors at the publisher and month levels. Figure 6 reveals the distribution of the effect of the GDPR on the number of trackers across these tracker categorizations.

By necessity, essential trackers experienced a significant decrease, with the DiD coefficient ($\beta = -1.192$, $p < 0.01$, 95% CI [$-1.961; -0.422$]) suggesting that the GDPR reduced the number of essential trackers by about one per publisher. Non-essential trackers also saw a significant reduction, with a DiD coefficient of $\beta = -2.757$ ($p < 0.05$, 95% CI [$-5.097; -0.417$]), indicating a decrease of about three per publisher.

By purpose, the results show that hosting trackers decreased significantly after the GDPR, as reflected by the DiD coefficient ($\beta = -0.305$, $p < 0.01$, 95% CI [$-0.485; -0.125$]). CDN trackers also experienced a significant reduction, with a DiD coefficient of $\beta = -0.792$ ($p < 0.01$, 95% CI [$-1.268; -0.316$]). Likewise, analytics trackers saw a substantial decrease, with a DiD coefficient of $\beta = -0.873$ ($p < 0.001$, 95% CI [$-1.271; -0.476$]), indicating a reduction of about

Figure 6: Distribution of the GDPR's Impact Across Categorizations of Trackers

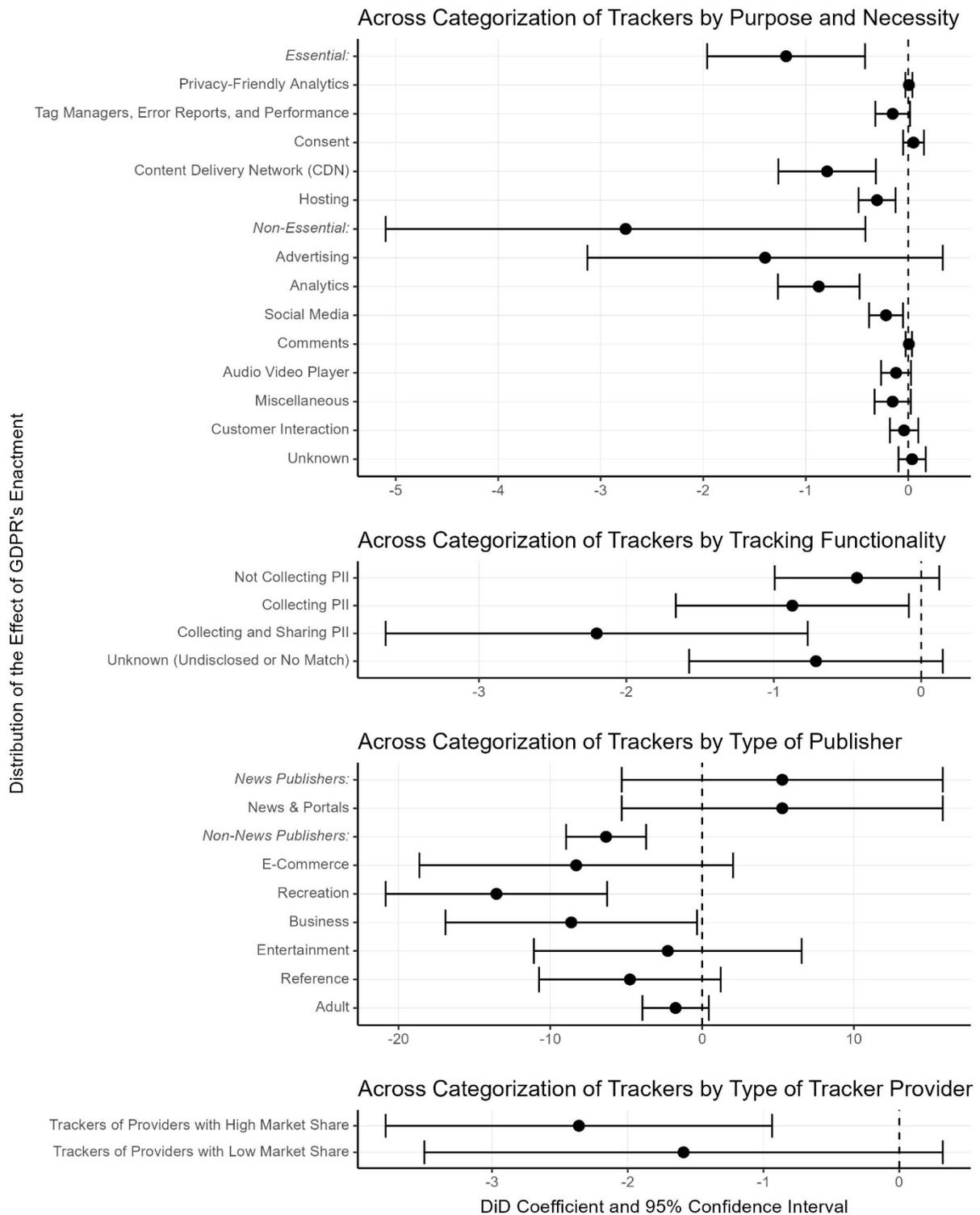

Notes: This figure shows the difference-in-differences coefficients (Treatment x PostGDPR) from OLS regressions and 95% confidence intervals, with the dependent variables being the number of trackers across different categorizations of trackers. Italicized labels represent grouped variables, where a broad category estimation (e.g., "Essential:") is followed by estimations for subcategories within that group (e.g., "Privacy-Friendly Analytics"). Except for the categorization by Type of Publisher, all models have N = 9,408 observations (294 publishers * 32 months). For the categorization by Type of Publisher, the number of observations varies: News Publishers have N = 928 (29 publishers * 32 months), Non-News Publishers have N = 6,328 (198 publishers * 32 months), and specific types of publishers are as follows: News & Portals (N = 928), E-Commerce (N = 480), Recreation (N = 224), Business (N = 2,048), Entertainment (N = 2,432), Reference (N = 736), and Adult (N = 2,528). The Governmental publisher type is excluded due to having only one publisher. All models include website instance and month fixed effects. Two-way standard errors are clustered at the publisher and month levels.

0.9 per publisher. Social media trackers were also significantly affected, with a DiD coefficient of $\beta = -0.217$ ($p < 0.05$, 95% CI [$-0.382;-0.052$]).

By tracking functionality, the results show that trackers that collect PII experienced a significant decrease, as indicated by the DiD coefficient ($\beta = -0.874, p < 0.05$, 95% CI [$-1.664;-0.084$]). Moreover, trackers that collect and share PII also saw a significant reduction after the GDPR, with a DiD coefficient of $\beta = -2.201$ ($p < 0.01$, 95% CI [$-3.633;-0.770$]).

By type of publisher, non-news publishers saw a significant decrease in the number of trackers after the GDPR, as indicated by a DiD coefficient of $\beta = -6.328$ ($p < 0.001$, 95% CI [$-8.963;-3.694$]). Within this broader publisher industry, the recreation types of publishers experienced a particularly significant reduction, with a DiD coefficient of $\beta = -13.551$ ($p < 0.05$, 95% CI [$-20.843;-6.260$]).

Lastly, by the size of the tracker provider, the number of trackers from tracker providers with a high market share significantly decreased after the GDPR, with a DiD coefficient of $\beta = -2.360$ ($p < 0.01$, 95% CI [$-3.783;-0.937$]).

### 5.3. *Insights from the Robustness Tests*

We conducted a series of robustness tests to ensure the reliability of our main analysis, addressing potential concerns about methodology and data limitations. Table 8 summarizes these robustness tests, with further details in the Web Appendices.

First, we performed alternative treatment assignments based on server location and user location to account for possible misclassifications of publishers into treatment and control groups. The results show that the GDPR reduced the number of trackers by 3.867 per publisher when using server location as the treatment assignment criterion and by 1.692 per publisher instance when using publisher designation and user location as the treatment assignment criterion. We estimated the monthly DiD coefficients and conducted placebo tests, confirming that the assumption of parallel trends likely holds.

Table 8: Summary of Robustness Tests

| Robustness Test | Fundamental Concern | Summary of Result | Web Appendix |
|---|---|---|---|
| Treatment assignment based on server location | Misclassification of publishers into treatment (EU) and control groups (non-EU) based on publisher's website traffic shares and top-level domain (TLD) | GDPR reduced the number of trackers by 3.867 per publisher with treatment assignment based on server location | 9.1 |
| Treatment assignment based on publisher designation and user location | Misclassification of publishers into treatment (EU) and control groups (non-EU) based on publisher's website traffic shares and top-level domain (TLD) | GDPR reduced the number of trackers by 1.692 per publisher instance with treatment assignment based on publisher designation and user location | 9.2 |
| Parallel trends assumption | Treatment and control groups do not follow same trends in the pre-treatment period (violation of parallel trends assumption) | Development of monthly DiD coefficients and placebo tests confirm the assumption likely holds | 9.3 |
| Spillover effects | GDPR spillovers affect control group (= violation of stable unit treatment value assumption) | GDPR reduced the number of trackers by 2.922 per publisher instance in the "cleanest" comparison between treatment (EU-located users visiting EU publishers) vs. control (US-located users visiting non-EU publishers) groups | 9.4 |
| Impact of GDPR on user behavior | GDPR inadvertently affects behavior of Ghostery users rather than publishers' use of trackers | No significant change in the number of Ghostery users (Chrome and Firefox) after GDPR | 9.5 |
| Anticipation and external shocks (early 2018) | Bias from publishers' early willingness to comply with GDPR (= anticipation assumption) or shocks unrelated to the GDPR (e.g., Cambridge Analytica) | GDPR reduced the number of trackers by 4.523 per publisher when removing the months of March, April, May and June 2018 | 9.10 |
| Skewness of the dependent variable | Skewness in the distribution of the number of trackers | GDPR reduced the logged number of trackers by 0.490 per publisher | 9.11 |
| Stability of publishers' website traffic shares | Misclassification of publishers due to potential changes in website traffic distributions over time when using a single point-in-time SimilarWeb data set | The average difference for EU publishers' website traffic shares was 11.08 pp, indicating stable website traffic distributions between public (single point in time) and proprietary (over time) SimilarWeb data sets | 9.12 |
| Generalized synthetic control method | Potential model misspecifications in the difference-in-differences (DiD) analysis | GDPR reduced the number of trackers by 5.303 per publisher | 9.14 |
| Unbalanced panel | Potential lack of representativeness due to excluding a large number of publishers from the balanced panel | GDPR reduced the number of trackers by 1.081 (treatment assignment based on TLD) and 0.825 (treatment assignment based on server location) per publisher in the unbalanced panel of 29,735 unique publishers | 9.15 |

Regarding spillover effects, we performed the "cleanest" comparison between EU-located users visiting EU publishers and US-located users visiting non-EU publishers. This robustness test demonstrated that the GDPR reduced the number of trackers by 2.922 per publisher instance, with minimal spillovers affecting the control group. Additionally, we tested whether the GDPR affected user behavior rather than publishers' tracking practices by examining the number of Ghostery users before and after the GDPR. The results indicated no significant change in Ghostery users, confirming that the reduction in trackers was due to publishers' tracking practices.

To address concerns about early compliance with the GDPR or shocks unrelated to the GDPR, such as the Cambridge Analytica scandal (Cadwalladr and Graham-Harrison 2018), we removed the months of March, April, May, and June 2018 from our main analysis. Even after doing so, the GDPR reduced the number of trackers by 4.523 per publisher, suggesting that anticipation or external shocks do not bias our results.

We also addressed the skewness in the distribution of the number of trackers by applying a log transformation to our dependent variable. The results showed that the GDPR reduced the logged number of trackers by 0.490 per publisher, confirming the robustness of our findings.

Another concern we addressed was the potential instability of publishers' website traffic shares over time, given that we relied on a single point-in-time SimilarWeb data set to designate publishers into treatment and control groups. By comparing public SimilarWeb data (collected in September 2021) with proprietary SimilarWeb data (collected from January 2018 to December 2019), we found that the average difference in EU traffic shares was 11.08 percentage points, indicating stable website traffic distributions over time.

We applied the generalized synthetic control (GSC) method to account for potential model misspecifications in the DiD analysis. This method constructs a counterfactual (i.e., control group) by matching treated and control units more accurately, relying on pre-treatment trends.

Unlike the DiD approach, GSC allows the algorithm to select the optimal control group. The GSC method's analysis showed that the GDPR reduced the number of trackers by 5.303 per publisher, further confirming the robustness of our findings.

Finally, we performed a robustness test using an unbalanced panel to ensure our results are representative for a larger number of publishers. While we use a balanced panel of 294 publishers in the main analysis to, among others, avoid panel attrition, the unbalanced panel includes 29,735 unique publishers, representing a 9,994% increase. It allows us to maximize the number of observations. We assigned treatment to publishers based on TLD and server location, and the results show that the GDPR reduced the number of trackers by 1.081 per publisher (treatment assignment based on TLD) and by 0.825 per publisher (treatment assignment based on server location). This robustness test also confirms the robustness of our findings.

## 6. Summary, Conclusions, and Implications

### 6.1. Summary of Results and Conclusions

Trackers are software that combines a specific purpose with tracking functionality. Publishers embed them into their websites to monitor user behavior, personalize content, or deliver targeted ads. However, trackers often collect and share user data across multiple publishers and advertisers. While they generate value for publishers by enhancing content and attracting users—monetized through targeted advertising—they raise substantial privacy concerns by processing users' personal data. Consequently, regulators have enacted laws like the EU's GDPR to enhance online privacy.

In this paper, we examined the impact of the GDPR on online tracking. By categorizing trackers in a manner aligned with the GDPR's objectives, we assessed both the intended and unintended consequences of the regulation. Our main findings and conclusions are as follows (see Table 9 and Table 10 for a summary):

**Table 9: Summary of Empirical Findings on Description of Online Trackers and Their Conclusions**

| Analysis | Summary of Findings | Conclusions |
|---|---|---|
| Average Effect | • Average number of trackers per publisher (~17)<br>• Most publishers use 1-10 trackers (Min= 1, Max = 111) | • Some publishers strongly rely on trackers for different purposes.<br>• Distribution of trackers is heavily right-skewed. |

*Differences across categorizations of online trackers*

| Analysis | Summary of Findings | Conclusions |
|---|---|---|
| Trackers by Necessity | • 27% essential trackers per publisher (~4)<br>• 73% non-essential trackers per publisher (~12) | • Publishers use three times as many non-essential than essential trackers.<br>• Users are exposed to privacy risks from non-essential trackers. |
| Trackers by Purpose | • Top essential trackers:<br>    ○ 66% content delivery (~3)<br>    ○ 17% tag managers (~1)<br>    ○ 13% hosting (~1)<br>• Top non-essential trackers:<br>    ○ 59% advertising (~7)<br>    ○ 23% analytics (~3)<br>    ○ 5% social media (~1)<br>• Among essential trackers:<br>    ○ 0.7% privacy-friendly analytics (<1) | • Advertising, analytics, and content delivery trackers are most often used.<br>• Publishers rarely use privacy-friendly analytics trackers. |
| Trackers by Functionality | • 11% of trackers do not collect personal data (~2)<br>• 66% of trackers collect personal data<br>    ○ 28% of those trackers do not share personal data (~3)<br>    ○ 72% of those trackers share personal data (~8) | • Most trackers are highly privacy invasive as they collect and share personal data. |
| Trackers by Type of Publisher | • 67% of trackers belong to news publishers (~30)<br>• 33% of trackers belong to non-news publishers (~15) | • News publishers use twice as many trackers as non-news publishers.<br>• News publishers rely on trackers to enhance and monetize their content through advertising. |
| Trackers by Size | • 50% of trackers belong to providers with a high market share (~8)<br>• 50% of trackers belong to providers with a low market share (~8) | • Publishers use a similar amount of trackers from tracker providers with a high or low market share.<br>• Across all trackers, our study does not find evidence for market concentration of large or small trackers. |

**Table 10: Summary of Empirical Findings of Impact of GDPR on Online Trackers and Their Conclusions**

| Analysis | Summary of Findings[A] | Conclusions |
|---|---|---|
| Average Effect | • Average reduction of trackers (~4) | • GDPR reaches its intended consequence and decreases trackers by 14.79% compared to expectations without GDPR.<br>• Although trackers increase over time across EU and non-EU publishers before and after the GDPR, the increase is much smaller for EU publishers. |
| | *Differences across categorizations of online trackers* | |
| Trackers by Necessity | • Average reduction of essential trackers (~1)<br>• Average reduction of non-essential trackers (~3) | • GDPR led to the unintended consequence of decreasing essential trackers.<br>• GDPR reached its intended consequence of decreasing non-essential trackers. |
| Trackers by Purpose | • Average reduction of essential trackers<br>  ○ Content Delivery (~1)<br>  ○ Hosting (~1)<br>  ○ Privacy-friendly analytics (~0)<br>• Average reduction of non-essential trackers<br>  ○ Analytics (~1)<br>  ○ Social Media (~1)<br>  ○ Advertising (~0) | • GDPR did not reach its intended consequence of decreasing advertising trackers and increasing privacy-friendly analytics trackers. |
| Trackers by Functionality | • Average reduction of trackers that do not collect personal data (~0)<br>• Average reduction of trackers that do collect personal data<br>  ○ Tracker does not share personal data (~1)<br>  ○ Tracker shares personal data (~2) | • GDPR achieved its intended consequence of decreasing highly invasive tracking. |
| Trackers by Type of Publisher | • Average reduction of trackers of news publishers (~0)<br>• Average reduction of trackers of non-news publishers (~6)<br>  ○ Recreation (~14)<br>  ○ Business (~9)<br>  ○ E-commerce (~0)<br>  ○ Entertainment (~0) | • GDPR reached its intended consequence and decreased tracking of non-news publishers.<br>• GDPR led to the unintended consequence of not decreasing trackers of news, e-commerce, and entertainment publishers. |
| Trackers by Size | • Reduction of average number of trackers of providers with high market share (~2)<br>• Reduction of average number of trackers of providers with low market share (~0) | • GDPR reached its intended consequence of decreasing trackers of high market share tracker providers and did not increase market concentration.<br>• GDPR reached the unintended consequence of not decreasing trackers of low market share tracker providers. |

Notes: A) The summary of findings refers to the average reduction of trackers per EU publisher.

First, publishers heavily rely on trackers for various purposes, with advertising, analytics, and content delivery being the most commonly used. Most trackers are highly privacy-invasive, as they collect and share personal data. Although the number of trackers increased for EU and non-EU publishers from before to after the GDPR, the increase was significantly smaller for EU publishers. Thus, the GDPR achieved its intended consequence by decreasing the number of trackers by 14.79% compared to expectations without the regulation.

Despite the GDPR's implementation, many trackers remain on publishers' websites, and thus, the infrastructure for collecting, retaining, and sharing data remains largely intact. Notably, the GDPR did not significantly reduce the number of advertising trackers and only marginally reduced the number of analytics trackers. This outcome favors publishers seeking to enhance their content, attract new users, and monetize them through ads. It also benefits advertisers aiming to reach their target audiences effectively and tracker providers monetizing collected data by offering enhanced tracking services or selling it to third parties.

Second, privacy concerns persist despite the continued presence of many trackers on publishers' websites—which provide value to the online advertising market and potentially to users. However, the introduction of the GDPR has allowed users to consent to or decline the processing of personal data, providing them with greater control over their personal information. It remains unclear whether all users will choose to consent to being tracked. Nonetheless, users now have the option to deny consent, which was not readily available before the GDPR. This option represents a significant benefit for users due to the regulation.

Third, advertisers face publishers that collect varying amounts of user data. While the GDPR did not achieve its intended consequence of decreasing advertising trackers, it reduced the use of highly invasive trackers. This reduction potentially leaves advertisers with less data from some publishers. There is substantial heterogeneity in tracker usage across publishers; most have between 1 and 10 trackers, but the distribution is heavily right-skewed, with some publishers using up to 111 trackers.

## 6.2. *Implications*

From the summary of our main findings and conclusions, we derive the following implications:

First, advertisers face considerable heterogeneity among publishers regarding the amount of user data collected. This heterogeneity implies that behavioral targeting may no longer function uniformly across all publishers. Advertisers can respond by reallocating their ad budgets toward publishers with more extensive user data and away from those with less data. Alternatively, they may invest in more privacy-preserving forms of advertising on publishers with limited data, such as contextual targeting or privacy-enhancing technologies like Google's Privacy Sandbox (Johnson 2024), to effectively reach their target audiences.

Second, it is unclear whether publishers that use less tracking will gain a competitive advantage through increased user engagement or will face disadvantages due to advertisers' reduced ability to employ behavioral targeting on their platforms. Advertisers might have to rely on alternative forms of advertising (e.g., contextual targeting) with these publishers, which could be less profitable than behavioral targeting. This uncertainty raises questions about the balance between enhancing user privacy and maintaining revenue streams for publishers.

Third, our study demonstrated that the GDPR achieved some of its intended consequences by reducing overall tracking by 14.79% and decreasing the use of highly privacy-invasive trackers that collect and share personal data. Nevertheless, even after the GDPR, publishers continue to use trackers heavily and are increasingly doing so. If regulators aim to reduce tracking further, further activities are necessary, maybe including even stricter enforcement of the existing rules.

Fourth, the GDPR potentially increased privacy for users, especially for those who exercised their right to decline consent for tracking. However, our results also indicate reduced content delivery trackers, which may have diminished user experience on publishers' websites. Users might receive less relevant content recommendations and advertisements with less personal data being tracked. The unintended consequence of the GDPR reducing essential trackers that do not collect or share personal data but deliver content (e.g., videos) negatively impacts users.

A potential way to preserve the advantages of tracking tools while addressing privacy concerns is to decouple content provision from the tracking functions. This decoupling would let users access enriched content without being subject to tracking. However, this approach would require someone—whether it's the publishers, advertisers, or even the users—to cover the cost of providing the tracker technology without its data-gathering capabilities. Essentially, such unbundling places the financial burden on one party to maintain the content-enhancing features while ensuring user privacy. Achieving a balance between privacy and functionality often involves trade-offs, and in this case, someone will have to bear that cost.

## 7.     Limitations and Future Research

Before concluding, we acknowledge three limitations of our study and suggest avenues for future research.

First, while the GDPR likely contributed to the reduction in trackers observed after its implementation, understanding the exact mechanisms behind this reduction is challenging. Other factors may have played a role, such as (i) Innovations by tracker providers that allowed publishers to comply with the GDPR while continuing to use trackers; (ii) Slow and inconsistent enforcement of the GDPR, which may have encouraged publishers to maintain or re-establish their use of trackers once they perceived lower risk of penalties; (iii) Introduction of consent mechanisms over time (e.g., consent trackers), helping publishers manage compliance without significantly affecting their business goals (Johnson et al. 2023; Lefrere et al. 2024). Future research could explore these mechanisms by engaging with publishers and tracker providers to understand how they adapted to the GDPR over time.

Second, the role of user consent in determining the number of trackers is complex. Studies like Demir et al. (2024) confirm that when users deny consent via GDPR-compliant cookie banners, publishers reduce the number of trackers deployed, contributing to lower tracking practices. However, other research indicates that publishers do not always honor user consent. Sanchez-Rola et al. (2019) and Bouhoula et al. (2024) show that some publishers continue to track users even after denying consent through cookie banners. Non-compliance can be intentional, as publishers may ignore users' choices to maintain monetization benefits despite risking GDPR fines, or unintentional if publishers are unaware that specific trackers remain active (Ghostery 2017; Müller-Tribbensee 2024). This ambiguity complicates the interpretation of tracker reductions as solely driven by user consent. Future research should investigate the extent of publisher compliance with consent mechanisms and how this affects tracking practices.

Third, while the WhoTracks.me dataset provides comprehensive metrics on publishers' tracking practices, it may not fully represent the general internet user population. The data comes from users who have installed privacy tools like Ghostery, who are typically more privacy-conscious and technologically savvy than average users. This self-selection could limit the generalizability of our findings. Privacy-conscious users may engage in different browsing behaviors and are more likely to have additional privacy-focused extensions installed, which might block trackers before Ghostery can detect them, leading to an underestimation of the number of trackers recorded. We mitigate this concern by using the unique number of trackers users encounter per publisher as our dependent variable. Since WhoTracks.me aggregates data from millions of users, as long as at least one user does not have additional privacy tools installed, the complete set of unique trackers a publisher uses can be captured.

Nevertheless, the potential lack of representativeness remains a limitation. Future research could incorporate more representative datasets that include a broader cross-section of users and compare results across different user groups to assess how varying levels of privacy awareness or cultural factors affect exposure to trackers. Such research could enhance the generalizability of findings related to publishers' tracking practices and the impact of privacy regulations like the GDPR.

# 9. Web Appendix

Table of Contents

### 9.1. Robustness Test Using Alternative Treatment Assignment Based on the Publisher's Server Location

In Section 4.3 of the main manuscript, we discussed different proxies to classify a publisher as an "EU" or "non-EU firm" under the GDPR. The approach in the main analysis uses a combination of the top-level domain (TLD) and traffic shares to designate publishers as EU or non-EU. However, as the GDPR applies to all firms processing EU user data, different proxies exist to identify if the publisher is an "EU" or a "non-EU firm" under the GDPR.

In this section, we explore an alternative treatment assignment based on the publisher's server location, which offers a different approach to identifying whether a publisher operates within the EU or outside of it. We assume publishers hosting their servers within the EU are more likely to be subject to the GDPR. By assigning treatment based on the publisher's server location, we test whether the results of the GDPR's impact on trackers are robust to this alternative treatment assignment.

To do so, we identify the server locations of the publishers in our sample. If the publisher's server is located inside the EU, we consider the publisher an "EU publisher". If the server is located outside the EU, we classify the publisher as a "non-EU publisher".

We show how the distribution of our sample's observations changes under this new treatment assignment framework (Table 11).

Table 11: Distribution of Observations (Monthly Publishers) Across Publisher Designation for the Sample with a Server Location Treatment Assignment

| Publisher's Designation | Number and Percentage of Observations |
|---|---|
| EU publisher[1] | 3,488 (37.07%) |
| Non-EU publisher[2] | 5,920 (62.93%) |
| ∑ | 9,408 (100.00%) |

[1]A publisher is designated as an "EU publisher" if the publisher's server is located inside the EU. [2]A publisher is designated as a "non-EU publisher" if the publisher's server is located outside the EU.
Notes: The cells in this table show the number and percentage of observations in our sample corresponding to each case. The cell belonging to the control group—where GDPR does not apply—is colored gray, and the cell belonging to the treatment group—where GDPR applies—is not colored. In total, 37% (N observations = 3,488) of all observations (N observations = 9,408) belong to the treatment group and 63% (N observations = 5,920) to the control group.

37.07% (N = 3,488) of all observations (N = 9,408) belong to the treatment group, and 62.93% (N = 5,920) to the control group.

We visualize the development of the average monthly number of trackers using this server location-based treatment assignment in Figure 7.

Figure 7: Development of the Average Monthly Number of Trackers with Treatment Assignment Based on Server Location

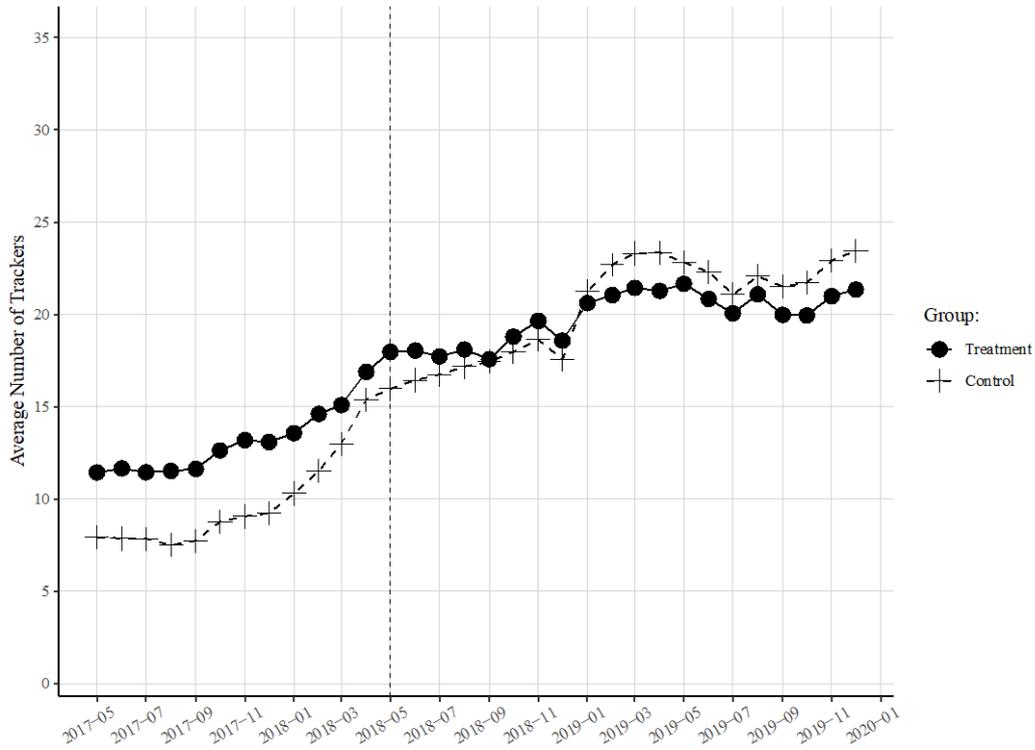

Figure 7 shows the trend in the average number of trackers used by the treatment and control groups over time. The number of trackers in the treatment and control groups increased over the observation period. Initially, the control group had fewer trackers than the treatment group, but the gap narrowed as time progressed. Around the time of the GDPR's enactment (May 2018), the treatment group exhibited a slight dip in the number of trackers, though the control group continued to increase.

We use the baseline OLS regression specified in Equation (1) to estimate the effect of the GDPR on the number of trackers. Table 12 outlines this result.

Table 12: Result of Difference-in-Differences Analysis for the Sample with a Server Location Treatment Assignment

| Dependent Variable: | Number of Trackers per Publisher and Month |
|---|---|
| Model: | (1) |
| Treatment x PostGDPR | -3.867** [-6.438; -1.296] |
| Publisher ID Fixed Effects | ✓ |
| Month ID Fixed Effects | ✓ |
| N Observations | 9,408 |
| $R^2$ | 0.745 |

Significance levels: * $p < 0.05$, ** $p < 0.01$, *** $p < 0.001$.
Two-way standard errors are clustered at the publisher and month levels; 95% confidence intervals are reported in brackets.
Notes: This table shows the difference-in-differences coefficient (Treatment x PostGDPR) from the OLS regression. We assign treatment to each publisher according to the publisher's server location (within the EU or outside the EU). Multiplying the number of publishers (N publishers = 294) and the number of months (T = 32 months) yields the number of observations (N observations = 9,408).

The DiD coefficient ($\beta$ = -3.867, $p < 0.01$, 95% CI [-6.438; -1.296]) is significantly negative in our DiD model presented in column (1). This result confirms that the GDPR reduced the number of trackers by about 3.867 trackers per publisher when using the publisher's server location as the treatment assignment criterion.

## 9.2. Robustness Test Using Alternative Treatment Assignment Based on the Publisher's Designation and User Location

In Section 4.3 of the main manuscript, we discussed different proxies to classify a publisher as an "EU" or "non-EU firm" under the GDPR. While our main analysis uses a combination of top-level domain (TLD) and traffic shares to designate publishers, we now explore an alternative treatment assignment based on the publisher's designation and the users' location visiting the publisher's website. This approach provides an additional layer of granularity by considering the origin of the audience accessing the publishers.

To do so, we use a secondary sample of the WhoTracks.me data, including information on the geographical location of users visiting the publishers (EU vs. US). This secondary sample consists of the same set of 294 publishers as used in the main analysis but only contains a single pre-treatment period (i.e., April 2018). We refer to a particular publisher–user base combination as a "publisher instance", such that, for each publisher in our secondary sample, there are two publisher instances: the publisher as experienced by EU users and the publisher as experienced by US users.

As before, we show how the distribution of our secondary sample's observations changes under this new treatment assignment framework (Table 13).

Table 13: Distribution of Observations (Monthly Publishers) Across Publisher Designation and User Location (EU vs. US)

| Publisher Designation | User Location | | ∑ |
|---|---|---|---|
| | EU | US | |
| EU publisher[1] | 1,386 (11.22%) | 1,386 (11.22%) | 2,772 (29.46%) |
| Non-EU publisher[2] | 4,788 (38.78%) | 4,788 (38.78%) | 9,576 (101.79%) |
| ∑ | 6,174 (50.00%) | 6,174 (50.00%) | 12,348 (131.25%) |

[1]A publisher is designated as an "EU publisher" if (1) the publisher uses an EU top-level domain (e.g., .de) or (2) the publisher receives more traffic from EU users than non-EU users. [2]A publisher is designated as a "non-EU publisher" if (1) the publisher uses a non-EU top-level domain (e.g., .com) and (2) the publisher receives more traffic from non-EU users than EU users.

Notes: The cells in this table show the number and the percentage of observations (corresponding to monthly publisher instances) in our sample corresponding to each case. The cell belonging to the control group—where GDPR does not apply—is colored gray, and the cells belonging to the treatment group—where GDPR applies—are not colored. In total, 61% (N observations = 7,560) of all observations (N observations = 12,348) belong to the treatment group, and 39% (N observations = 4,788) belong to the control group.

As Table 13 shows, 61% (N = 7,560) of all observations (N = 12,348) belong to the treatment group, and 39% (N = 4,788) to the control group.

Figure 8 shows the development of the average number of trackers over time in this secondary sample.

Figure 8: Development of the Average Monthly Number of Trackers in the Secondary Sample with Treatment Assignment Based on Publisher Designation and User Location (EU vs. US)

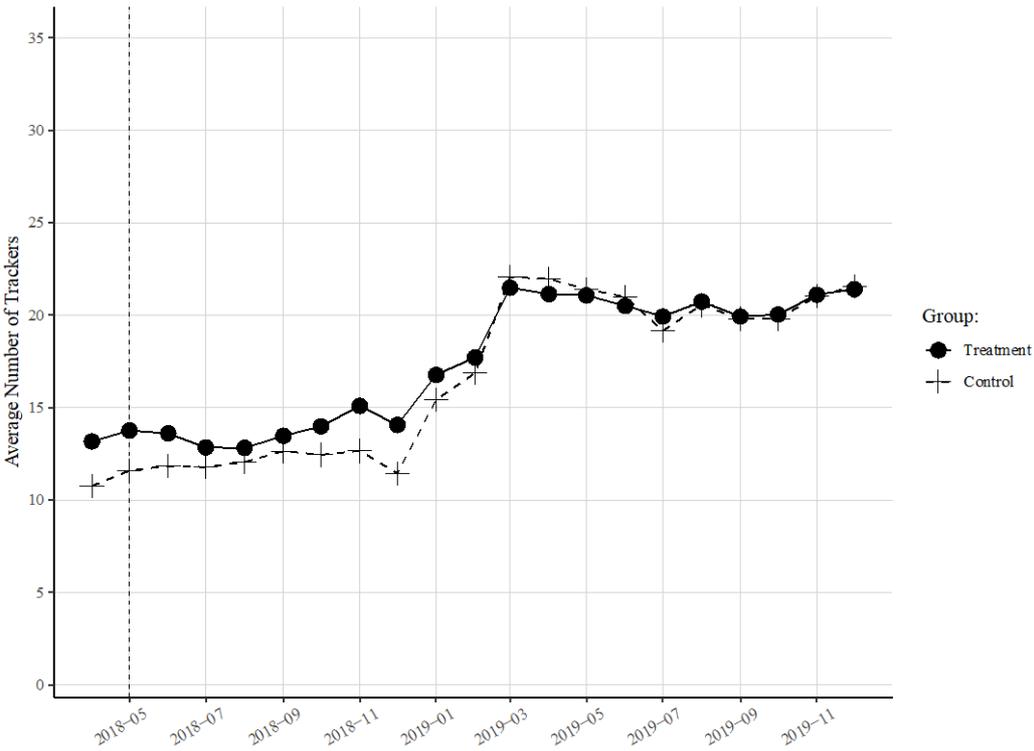

Figure 8 suggests an increasing trend in the number of trackers in the treatment and control groups. We further observe that the number of trackers in the treatment group is higher than in the control group each month, except in March 2019. Around the time of the GDPR's enactment (May 2018), the number of trackers in the treatment group slightly decreased, whereas that in the control group remained stable. However, by the end of 2018, the number of trackers in the treatment group was at about the same level as before the GDPR. Moreover, both groups increased the number of trackers from December 2018 onward, reaching a new level in March 2019. The number of trackers decreased slightly for both groups from that point onward.

We use the baseline OLS regression specified in Equation (1) to estimate the effect of the GDPR on the number of trackers under this new treatment assignment. Table 14 presents the results, comparing the secondary sample with a single pre-treatment period (i.e., April 2018) and the main sample, where we average all pre-treatment periods (May 2017 – April 2018) into a single pre-treatment period (i.e., April 2018) to make it comparable to the secondary sample.

Table 14: Result of Difference-in-Differences Analysis for the Number of Trackers of a Secondary Sample with Treatment Assignment Based on Publisher Designation and User Location (EU vs. US)

| Dependent Variable: | Number of Trackers per Publisher Instance and Month | Number of Trackers per Publisher and Month |
|---|---|---|
| Model: | (1) | (2) |
| Treatment x PostGDPR | -1.692*** [-2.259; -1.125] | -3.949*** [-5.047; -2.851] |
| Publisher Instance ID Fixed Effects | ✓ | ✗ |
| Publisher ID Fixed Effects | ✗ | ✓ |
| Month ID Fixed Effects | ✓ | ✓ |
| N Observations | 12,348 | 6,174 |
| $R^2$ | 0.779 | 0.833 |

Significance levels: * $p < 0.05$, ** $p < 0.01$, *** $p < 0.001$.
Two-way standard errors are clustered at the publisher instance and month levels in Model (1); Two-way standard errors are clustered at the publisher and month levels in Model (2); 95% confidence intervals are reported in brackets. Notes: This table shows the difference-in-differences coefficient (Treatment x PostGDPR) from the OLS regressions. In model (1), we assign treatment to each publisher instance according to the publisher's designation (EU or non-EU) and the users who visited it (EU vs. US). Multiplying the number of publishers (N publishers = 294), the number of publisher instances (2) and the number of months (T = 21 months) yields the number of observations (N observations = 12,348) in Model (1). In model (2), we assign treatment to each publisher according to the publisher's designation (EU or non-EU). Multiplying the number of publishers (N publishers = 294) and the number of months (T = 21 months) yields the number of observations (N observations = 6,174) in Model (2).

In Model (1), the significantly negative DiD coefficient ($\beta$ = -1.692, $p < 0.01$, 95% CI [-2.259; -1.125]) indicates that the GDPR reduced the number of trackers by about 1.692 trackers per publisher instance in the secondary sample. In model (2), the significantly negative DiD coefficient ($\beta$ = -3.949, $p < 0.001$, 95% CI [-5.047; -2.851]) remains consistent with our main findings when using the same treatment assignment but averaged across all pre-treatment periods to make it comparable to the secondary sample. These results confirm that our main findings are robust when comparing different samples and the treatment assignments.

### 9.3. *Robustness Test Regarding the Parallel Trends Assumption*

A fundamental assumption of the Difference-in-Differences (DiD) analysis is that the treatment and control groups would have followed parallel trends without the treatment—in this case, the GDPR's enactment. We conduct two tests to verify whether this assumption likely holds in our analysis: estimating monthly DiD coefficients (i.e., dynamic treatment effects) and performing a placebo test.

First, we estimate the monthly DiD coefficients to observe how the number of trackers evolved in the treatment and control groups before and after the GDPR.

Figure 9 shows these coefficients, with April 2018—the month immediately preceding the GDPR—used as the reference month. The model includes publisher- and month-fixed effects, and we cluster two-way standard errors at the publisher and month levels.

Figure 9: Development of the Monthly Difference-in-Differences Coefficients for the Number of Trackers

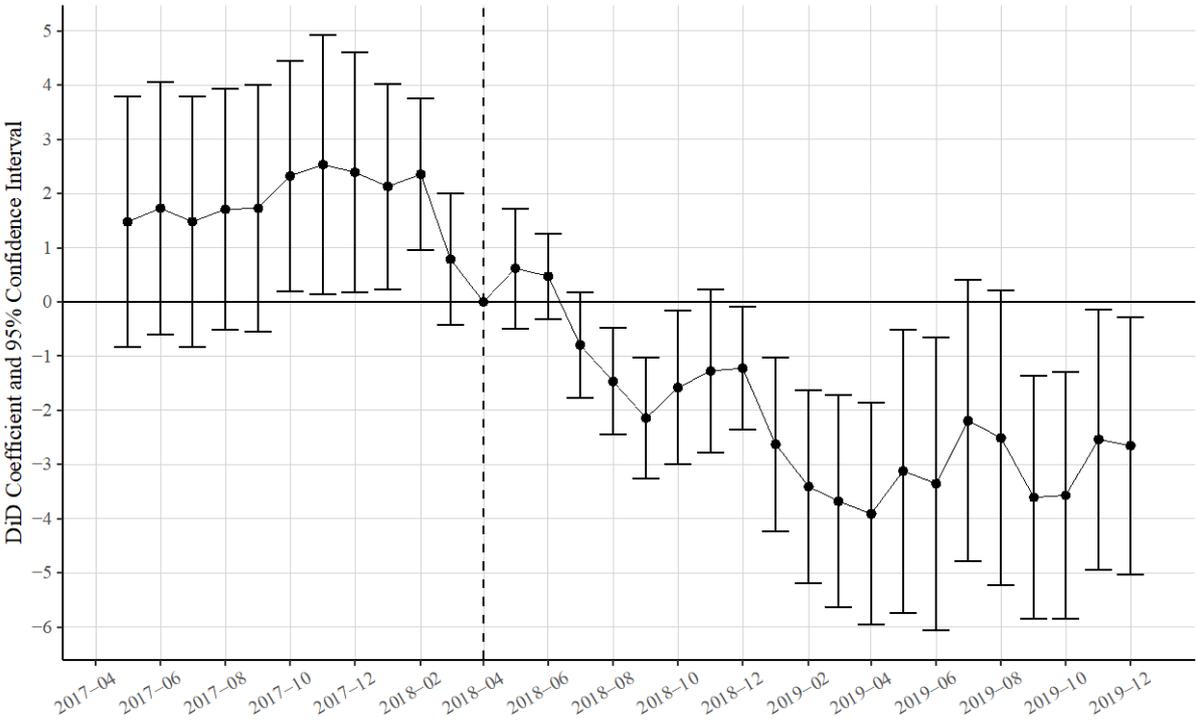

Notes: This figure shows the monthly difference−in−differences coefficients (Treatment x PostGDPR) from the OLS regression. We assign treatment to each publisher according to the publisher's designation (EU or non−EU). The model includes publisher and month fixed effects. Two−way standard errors are clustered at the publisher and month levels. The first month (April 2018) before GDPR's enactment serves as a reference month in the estimation, so its coefficient is zero.

Before May 2018, 6 out of 11 pre-treatment coefficients are insignificant, suggesting the parallel trends assumption likely holds. The few significant coefficients in the pre-GDPR period could be due to chance. After May 2018, the coefficients become significantly negative, reflecting a reduction in the number of trackers in the treatment group relative to the control group.

Second, we conduct a placebo test to assess the parallel trends assumption further. For this test, we pretend the GDPR's enactment (i.e., placebo GDPR treatment) occurred each month from June 2017 to April 2018 and observe whether any significant treatment effect arose before the GDPR (May 2017 – April 2018). Table 15 presents the results of these placebo regressions.

Table 15: Result of Placebo Difference-in-Differences Analysis for the Number of Trackers

| Dependent Variable: | Number of Trackers per Publisher and Month | | | | | | | | | | |
|---|---|---|---|---|---|---|---|---|---|---|---|
| Model: | (1) | (2) | (3) | (4) | (5) | (6) | (7) | (8) | (9) | (10) | (11) |
| Placebo GDPR Treatment in: | 2017-06 | 2017-07 | 2017-08 | 2017-09 | 2017-10 | 2017-11 | 2017-12 | 2018-01 | 2018-02 | 2018-03 | 2018-04 |
| Treatment x PostPretend | 0.265 [-0.301; 0.832] | 0.141 [-0.889; 1.170] | 0.211 [-0.993; 1.414] | 0.182 [-1.157; 1.522] | 0.164 [-1.322; 1.650] | -0.042 [-1.627; 1.544] | -0.321 [-2.109; 1.466] | -0.603 [-2.659; 1.453] | -0.898 [-3.217; 1.422] | -1.592. [-3.473; 0.289] | -1.877** [-2.960; -0.793] |
| Publisher ID Fixed Effects | ✓ | ✓ | ✓ | ✓ | ✓ | ✓ | ✓ | ✓ | ✓ | ✓ | ✓ |
| Month ID Fixed Effects | ✓ | ✓ | ✓ | ✓ | ✓ | ✓ | ✓ | ✓ | ✓ | ✓ | ✓ |
| N Observations | 3,528 | 3,528 | 3,528 | 3,528 | 3,528 | 3,528 | 3,528 | 3,528 | 3,528 | 3,528 | 3,528 |
| $R^2$ | 0.900 | 0.900 | 0.900 | 0.900 | 0.900 | 0.900 | 0.900 | 0.900 | 0.900 | 0.900 | 0.900 |

Significance levels: * $p < 0.05$, ** $p < 0.01$, *** $p < 0.001$.
Two-way standard errors are clustered at the publisher and month levels; 95% confidence intervals are reported in brackets.
Notes: This table shows the difference-in-differences coefficient (Treatment x PostPretend) from each OLS regression used for the placebo test. We use observations before GDPR's enactment (May 2017-April 2018) and assign treatment to each publisher using the publisher's designation (EU or non-EU). Multiplying the number of publisher (N publishers = 294), and the number of months before GDPR's enactment (T = 12 months) yields the number of observations (N observations = 3,528).

In 10 out of 11 months before the GDPR, we do not observe a significant placebo effect, which supports the validity of the parallel trends assumption. However, in April 2018, we find a significant negative effect ($\beta$ = -1.877, $p < 0.01$, 95% CI [-2.960; -0.793]), likely indicating that some publishers began adjusting their tracker usage shortly before the GDPR's official enactment.

In summary, the monthly DiD coefficient estimates and the placebo test results suggest that the parallel trends assumption is likely to hold true in our main analysis. The presence of a single significant placebo effect in April 2018 could be attributed to anticipation of the GDPR's enactment and does not undermine the validity of our analysis, especially given the insignificant effects in the months preceding it.

### 9.4. Robustness Test Regarding the Spillover Assumption

As discussed in Section 4.4 in the main manuscript, a fundamental assumption for our DiD analysis is the stable unit treatment value assumption (SUTVA). This assumption includes the expectation that no spillovers exist between the treatment and control groups, meaning that the GDPR should not affect publishers in the control group. A concern is that some publishers in the control group may have altered their tracking practices following the GDPR (e.g., to avoid a GDPR fine) to "comply voluntarily", potentially contaminating our control group. In this robustness test, we test for the spillovers and examine their potential magnitude.

An important point to consider is the potential effect of spillovers on our results. If strong spillovers were present, we expect the treatment and control groups to become more similar, which would lower the impact of the GDPR on the number of trackers rather than increase it. In this case, any differences we detect may underestimate the true impact of the GDPR. The presence of spillovers would mean that the GDPR's impact is larger than our results suggest, as spillovers would blur the differences between the treatment and control groups.

We build upon the secondary sample introduced in the Web Appendix 9.2 to test for spillovers. This secondary sample of the WhoTracks.me data includes 294 publishers as in our main analysis but is limited to a single pre-treatment period (April 2018). We identify two "publisher instances" for each publisher: one targeted to EU users and the other to US users. This sample allows us to assign treatment based on the publisher's designation (EU vs. non-EU) and the geographical location of users (EU vs. US) visiting the publisher, allowing us to test for spillovers and examine their potential magnitude.

We run our baseline OLS regression specification from Equation (1) on this secondary sample to compare how the GDPR's effects differ across various user–publisher combinations presented in Table 13. We consistently use Cell 4 (US users visiting non-EU publishers) as the control group while redefining the treatment group for each of the following three comparisons: (1) EU users visiting EU publishers (Cell 1) vs. US users visiting non-EU publishers (Cell 4), (2) EU users visiting non-EU publishers (Cell 3) vs. US users visiting non-EU publishers (Cell 4), and (3) US users visiting EU publishers (Cell 2) vs. US users visiting non-EU publishers (Cell 4). We present these results in Table 16.

Table 16: Result of Difference-in-Differences Analysis for the Number of Trackers Between Different Definitions of the Treatment and Control Groups

| Dependent Variable: | Number of Trackers per Publisher Instance and Month | | |
|---|---|---|---|
| | (1) | (2) | (3) |
| Model: | | | |
| | Cell 1 vs. Cell 4 | Cell 3 vs. Cell 4 | Cell 2 vs. Cell 4 |
| Illustration | 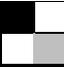 | 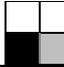 | 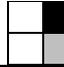 |
| Treatment x PostGDPR | -2.922*** | -1.907*** | 0.281 |
| | [-4.103; -1.741] | [-2.586; -1.229] | [-0.543; 1.105] |
| Publisher Instance ID Fixed Effects | ✓ | ✓ | ✓ |
| Month ID Fixed Effects | ✓ | ✓ | ✓ |
| N Observations | 6,174 | 9,576 | 6,174 |
| $R^2$ | 0.792 | 0.780 | 0.753 |

Significance levels: * $p < 0.05$, ** $p < 0.01$, *** $p < 0.001$.
Two-way standard errors are clustered at the publisher instance and month levels; 95% confidence intervals are reported in brackets.
Notes: This table shows the difference-in-differences coefficients (Treatment x PostGDPR) from the OLS regressions between different cells in our treatment assignment framework. We assign treatment to each publisher instance according to the publisher's designation (EU or non-EU) and the users who visited it (EU vs. US). Multiplying the number of publisher instances (N publisher instances) per cell(s) in our sample and the number of months (T = 21 months) yields the number of observations (N observations) for each model.

In column (1), we compare Cell 1 (EU users visiting EU publishers) to Cell 4 (US users visiting non-EU publishers), which we consider the "cleanest" comparison between the treatment and the control groups because this comparison is least likely to experience spillovers, as these groups have minimal interaction. We find a significantly negative effect of the GDPR on the number of trackers ($\beta$ = -2.922, $p < 0.001$, 95% CI [-4.103; -1.741]).

In column (2), we compare Cell 3 (EU users visiting non-EU publishers) to Cell 4, finding another significantly negative effect of the GDPR ($\beta$ = -1.907, $p < 0.001$, 95% CI [-2.586; -1.229]).

Finally, in column (3), we compare Cell 2 (US users visiting EU publishers) to Cell 4 and observe no significant effect of the GDPR on the number of trackers ($\beta$ = 0.281, p = 0.494, 95% CI [-0.543; 1.105]).

The results from this robustness test confirm the validity of our main analysis. We observe the strongest impact of the GDPR in comparing EU users visiting EU publishers (treatment group) vs. US users visiting non-EU publishers (control group), which is least likely to be contaminated by spillovers. The diminishing effects of the GDPR observed in two other comparisons suggest that the GDPR primarily affects interactions between EU users and EU publishers.

Notably, the results of the "cleanest" comparison (EU users visiting EU publishers vs. US users visiting non-EU publishers) are very close to those of the main analysis, with the GDPR effect of -2.922 trackers in this robustness test compared to -3.949 trackers in the main analysis (a difference of $-1.027$ trackers between the two DiD coefficients, or 26.01%). This similarity suggests that spillovers, if present, are limited and do not substantially bias our results. The small magnitude of the difference (26.01%) between the "cleanest" comparison and the main analysis indicates that spillovers are unlikely to distort the results meaningfully.

Further, it is essential to note that spillovers, if present, would bias the main results downward, meaning that our estimates of the GDPR's effect are likely conservative. In other words, any spillovers would make the treatment and control groups more similar, reducing the observed effect of the GDPR. The fact that we still observe significant effects, even with the possibility of spillovers, suggests that the true impact of the GDPR may be even larger than reported. Therefore, our main findings remain robust, and spillovers, if present, would only reinforce the GDPR's impact on the publishers' use of trackers.

### 9.5.  *Robustness Test Regarding the Impact of the GDPR on User Behavior*

A potential concern in using the WhoTracks.me dataset is that the GDPR may have influenced user behavior, leading users to visit different publishers with varying numbers of trackers. This change in user behavior could confound our analysis, as we might unintentionally measure the effect of the GDPR on user behavior rather than its impact on a publisher's tracker usage. Another related concern is the potential change in the number and composition of Ghostery users during the observation period, which could affect the data collection process of WhoTracks.me.

To address these concerns, we analyzed the number of Ghostery browser extension users over time to determine whether there were any significant changes after the GDPR. WhoTracks.me collected data from three sources: (1) the Cliqz browser, (2) the Cliqz browser extension for Firefox, and (3) the Ghostery browser extension compatible with Firefox, Safari, Chrome, Opera, and Edge browsers (WhoTracks.me Privacy Team, 2017b). Since Cliqz discontinued its browser and Firefox extension, we focused on changes in the number of Ghostery browser extension users for Firefox and Chrome browsers.

We utilized historical information from the Internet Archive's Wayback Machine, which captures snapshots of web pages over time, to determine the number of Ghostery extension users for Firefox and Chrome browsers. Specifically, we examined weekly captures of the Ghostery browser extension product pages on the Chrome and Firefox web stores from the respective observation periods.

We used different observation periods for Chrome and Firefox extensions for this robustness test. For the Chrome extension, Internet Archive's data is only available from April 2018 onward, so we use the period from April 2018 to December 2019. For the Firefox extension, Internet Archive's data is available before April 2018, allowing us to use the entire period from May 2017 to December 2019, which aligns with the main sample's observation period. Figure 10 illustrates the development of the number of Ghostery extension users over time.

Figure 10: Development of the Number of Ghostery Extension Users for Chrome and Firefox Browsers per Wayback Machine Capture Date

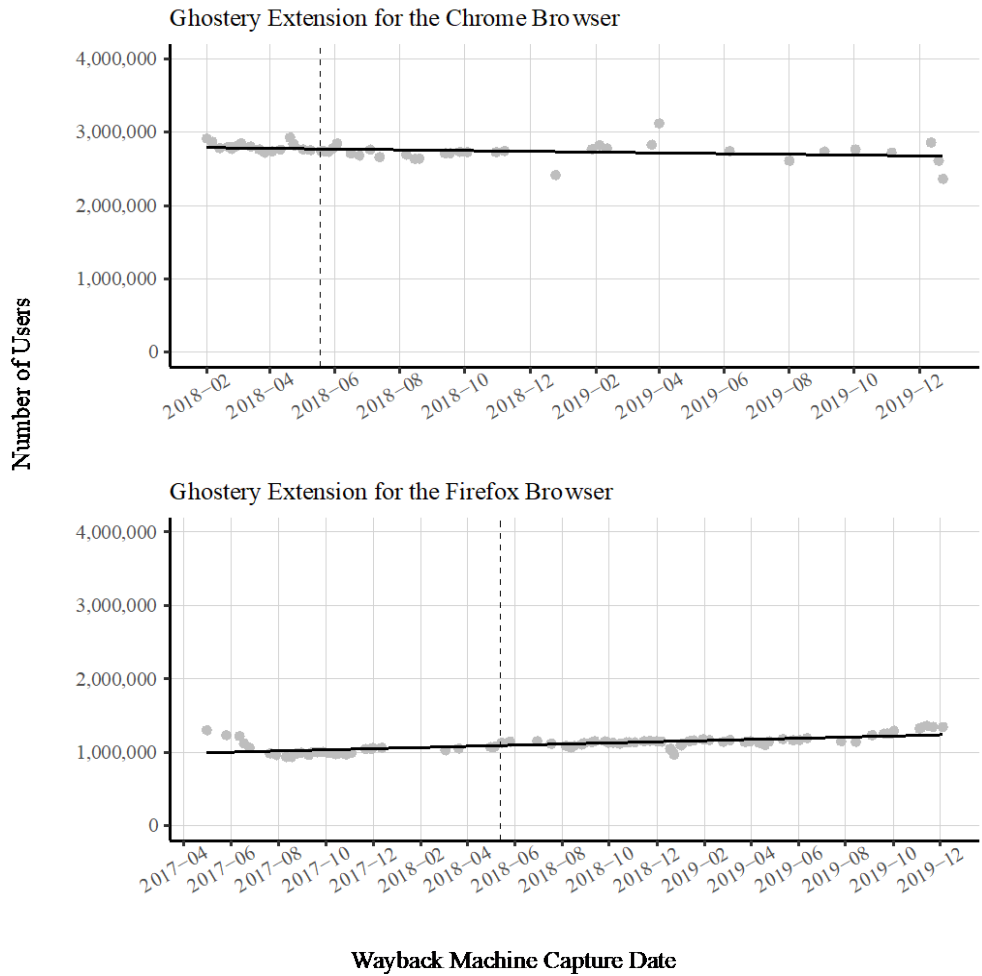

To evaluate the potential impact of the GDPR on the number of users of the Ghostery browser extension (for Chrome and Firefox), we specify the following OLS regression:

$$Y_t = \alpha + \beta PostGDPR_t + \tau Time_t + \epsilon_t \qquad (5)$$

In Equation (5), $Y_t$ represents the dependent variable, which is the number of users at a Wayback Machine capture date $t$. The intercept term $\alpha$ captures the baseline (average) number of users before considering other factors, corresponding to when $PostGDPR_t = 0$ and $Time_t = 0$. The coefficient $\beta$ associated with $PostGDPR_t$ is our coefficient of interest as it captures the difference in the number of users after the GDPR's enactment ($PostGDPR_t = 1$) compared to before ($PostGDPR_t = 0$). The term $\tau$ associated with $Time_t$ represents the linear time trend, which accounts for any gradual changes in the number of users over time, independent of the GDPR.

This effect is modeled based on the number of days elapsed since April 1$^{st}$, 2018. $\epsilon_t$ represents the error term. In summary, Equation (5) enables the assessment of the difference in the number of Ghostery extension users after the GDPR's enactment while accounting for a linear time trend.

We use the OLS regression specification from Equation (5) to calculate the coefficient for the number of Ghostery extension users and present the result in Table 17.

Table 17: Result of Regressions for the Number of Ghostery Extension Users

| Dependent Variable: | Number of Users per Wayback Machine Capture Date | | | |
|---|---|---|---|---|
| Model: | (1) | (2) | (3) | (4) |
| | Ghostery Extension for Chrome Browser | | Ghostery Extension for Firefox Browser | |
| Constant | 2,803,998.353*** [2,751,211.106; 2,856,785.600] | 2,803,714.356*** [2,750,660.244; 2,856,768.468] | 1,040,287.345*** [1,010,040.541; 1,070,534.148] | 1,083,615.575*** [1,051,311.216; 1,115,919.934] |
| PostGDPR | -82,535.697* [-147,856.582; -17,214.812] | -61,285.093 [-147,339.716; 24,769.529] | 122,743.211*** [85,244.041; 160,242.380] | -9,504.062 [-73,704.139; 54,696.016] |
| Time | | -76.634 [-277.295; 124.027] | | 272.386*** [159.290; 385.483] |
| N Observations | 49 | 49 | 83 | 83 |
| R$^2$ | 0.121 | 0.132 | 0.344 | 0.490 |
| Adjusted R$^2$ | 0.102 | 0.094 | 0.336 | 0.477 |

Significance levels: * $p < 0.05$, ** $p < 0.01$, *** $p < 0.001$.
95% confidence intervals are reported in brackets.
Notes: This table displays the PostGDPR coefficients obtained from OLS regressions. PostGDPR coefficients represent the difference in the number of users of the Ghostery browser extension after the GDPR's enactment. Models (1) and (2) have 49 observations based on weekly captures of Google Chrome's Ghostery product listing page from the Internet Archive's Wayback Machine. Models (3) and (4) have 83 observations, using weekly captures from Firefox's Ghostery product listing page.

In our baseline model presented in column (2), the $\beta$ coefficient ($\beta PostGDPR_t$) is not significantly negative ($\beta = -61,285.093$, $p = 0.158$, 95% CI [-147,339.716; 24,769.529]). Similarly, in our baseline model presented in column (4), the coefficient is not significantly negative ($\beta = -9,504.062$, $p = 0.769$, 95% CI [-73,704.139; 54,696.016]). However, in the model presented in column (3), the $\beta$ coefficient ($\beta PostGDPR_t$) is significantly positive ($\beta = 122,743.211$, $p < 0.001$, 95% CI [85,244.041; 160,242.380]). It is important to note that this model does not account for the linear time trend, similar to model (1) for Chrome, which shows a significant decline in users.

These results confirm that the number of users of the Ghostery browser extension (for Chrome and Firefox) remained unchanged after the GDPR. The finding suggests that the GDPR did not impact the number of Ghostery extension users. Therefore, we can conclude that we are measuring the impact of the GDPR on a publisher's usage of trackers rather than its effect on user behavior.

### 9.6.    *Overview of the Raw Data Available from WhoTracks.me*

As mentioned in Section 4.1 in the main manuscript, we used publicly available data from WhoTracks.me as our primary data source. In this section, we outline the raw data available from WhoTracks.me.

WhoTracks.me is a tool that collects (anonymous and voluntarily shared) data from users of Ghostery and Cliqz—browsing tools that protect user privacy—to improve the anti-tracking algorithms underlying those tools (Karaj et al. 2018b; WhoTracks.me n.d.). In addition to this function, WhoTracks.me uses the data it collects to provide a public "census of trackers across the web".

More specifically, WhoTracks.me has access to data, on a user level, of all trackers that a particular (anonymized) user encounters while visiting a specific publisher (see Web Appendix 9.7 for details). WhoTracks.me obtained these data from users of (1) the Cliqz browser, (2) the Cliqz browser extension for Firefox, and (3) the Ghostery browser extension for the Firefox, Safari, Chrome, Opera, and Edge browsers, who opt-in to voluntarily share data with WhoTracks.me (Greif 2017). Cliqz (browser and extension) users are based primarily in Germany, whereas users of the Ghostery browser extension are distributed worldwide (Karaj et al. 2018b; WhoTracks.me n.d.).

Notably, because it obtains data from individual users, WhoTracks.me (unlike other tracker detection mechanisms such as crawlers) has access to information about tracking activity on the sub-pages of each publisher's website, as well as tracking activity behind cookie banners (i.e., after the user expresses their consent) and on websites that require users to log in.

Based on the data it collects, WhoTracks.me publicizes, every month, aggregated publisher-level tracking metrics. Though WhoTracks.me has access to data for half a million publishers, it only releases data about a much smaller set of "top" publishers, namely, those browsed by a large enough group of Cliqz and Ghostery users during the previous month (Karaj et al. 2018a).

During the period at the focus of this study (May 2017–December 2019), WhoTracks.me publicized publisher-level metrics for about 8,334 publishers per month, around the $10^{th}$ day of each month. These data were publicly available on GitHub in CSV format, where each set of CSV files contained data corresponding to the previous month. The CSV files available on GitHub contained the following information for each publisher, among other details: (i) publisher industry; (ii) list of the exact trackers' users encountered. In a separate tracker database file, WhoTracks.me also publicizes more detailed information for each tracker, among other details: (v) tracker category; (vi) tracker provider; (vii) tracker domains.

WhoTracks.me also publicized separate sets of publisher metrics obtained from users corresponding to different regions, including the US and the EU. Thus, for a given publisher, it was possible to observe the tracking metrics obtained from US users and those obtained from EU users separately.

Though we do not have data on the precise numbers of users who contributed the data from which WhoTracks.me derived the tracking measurements publicized during our observation period, we can provide a rough estimate of the overall quantity of data that WhoTracks.me had at its disposal during this time. Specifically, WhoTracks.me has stated that from May 2017 to December 2017, it received data corresponding to 100 million loaded webpages per month on average, a number that increased to over 300 million loaded webpages per month from April 2018 (Karaj et al. 2018b; WhoTracks.me n.d.). About 5 million users visiting more than half a million publishers contributed this data. In blog posts from April, May, and June 2018, the WhoTracks.me team stated that they had received 360, 340, and 370 million loaded webpages during March, April, and May 2018, respectively (Karaj et al. 2018b; WhoTracks.me n.d.). Suppose we assume similarly loaded webpages throughout our observation period, taking an average across the latter three months. In that case, we can assume that WhoTracks.me received, on average, 357 million loaded web pages per month in our observation period from May 2017 to December 2019: a total of 11.4 billion loaded web pages over 32 months.

We acknowledge that the GDPR may have affected how WhoTracks.me collected data, thereby influencing the reliability of this data source. However, we suggest that such an effect is unlikely to have occurred. First, even before the GDPR, WhoTracks.me operated in a manner compatible with the GDPR requirements. Specifically, WhoTracks.me obtained users' explicit consent to collect their data, and it did not change its tracking method after the GDPR.

Moreover, we did not find evidence that the number and composition of Ghostery users tracked by WhoTracks.me changed significantly throughout our observation period (see Web Appendix 9.5).

Additionally, we collected information about the historical changes on GitHub (i.e., "commits") to ensure that no significant updates in the tracker database affected how the WhoTracks.me team collected and calculated its online tracking metrics during our observation period. Likewise, we have no reason to believe that users disadopted WhoTracks.me products due to GDPR. Conversations with the WhoTracks.me team that releases the data further support our assumption that the publicly available data sets on WhoTracks.me are reliable and of high quality.

### 9.7.    *Additional Background on WhoTracks.me's Data Collection Process*

To further clarify the nature of the data obtained from WhoTracks.me, we provide the following illustration of the publisher's data-generating process. Suppose a Chrome browser user installs the Ghostery browser extension from the Chrome Web Store. She does so to (1) stop seeing ads/pop-ups on webpages for a cleaner browsing experience (i.e., ad-blocking), (2) block tracking technologies on webpages to increase her online privacy (i.e., anti-tracking), or (3) monitor tracking technologies, e.g., across firm-owned webpages to verify that they function correctly (i.e., monitoring tracking technologies).

After installing the Ghostery extension, the user configures tracker categories (e.g., advertising, analytics, social media) that the extension should block/detect and decides to share her anonymous data with WhoTracks.me. Then, she begins to browse the web. The first webpage she visits is example.com/home, the homepage of the example.com publisher. The Ghostery extension scans the Hypertext Transfer Protocol (HTTP) request that this visit creates. Ghostery detects, among other things, the requested URLs associated with the example.com/homepage and separates (1) URLs belonging to the requested publisher's server (i.e., a first-party domain, example.com) from (2) URLs belonging to servers that are external to the publisher (i.e., third-party domains like doubleclick.net or google-analytics.com). Essentially, Ghostery intercepts communication between a webpage the user visits and its external resources (e.g., ads, video, images, code).

Next, Ghostery uses the third-party domains it has detected to identify the trackers on the webpage by matching their domains with its database of domains that trackers use. If Ghostery detects multiple domains of the same tracker on the webpage (e.g., DoubleClick can use doubleclick.net, invitemedia.com, and 2mdn.net domains), Ghostery counts that tracker once. Depending on the user configuration, Ghostery blocks advertising trackers (e.g., DoubleClick) on the webpage but records and transmits user-observed trackers and their behavior (e.g., tracking via cookies) to WhoTracks.me—as the process runs alongside Ghostery's blocking (Karaj et al. 2018a). Because another user-visited webpage could include private information in its URL (e.g., example.com/firstname-lastname), the data delivered to WhoTracks.me are anonymized. If a large enough group of users visits the webpages of the example.com publisher (i.e., the example.com publisher qualifies as a "top publisher" according to WhoTracks.me), WhoTracks.me publicly releases tracking metrics of the example.com publisher.

The user can install other ad-blocking or anti-tracking browser extensions (e.g., Adblock Plus, uBlock Origin) in addition to Ghostery. In such a case, those extensions might block third-party domains—and "piggybacking" domains of those third-party domains—before Ghostery can detect them on users' browsed webpages. The WhoTracks.me team has explained that such a situation would lead to an underestimation of the average number of trackers per webpage on publishers that the user visits (WhoTracks.me Privacy Team, 2018a).

### 9.8. Description of the Sample Construction Process

This section describes how we constructed the sample of 294 publishers for our main analysis. For context, WhoTracks.me releases two types of data sets: the "global" and the "EU/US" samples.

The global sample contains data from May 2017 until December 2019, covering 32 months. It includes detailed tracking information for websites (referred to as publishers), but it does not provide any data about the location of the users visiting these publishers. In the raw global data, WhoTracks.me released an average of 8,334 publishers per month across this period, with the number of publishers increasing from 3,645 in May 2017 to 15,004 by May 2018. The panel of publishers in the raw data is unbalanced, as the number of publishers released each month varies, with a maximum of 17,987 publishers in September 2018. After filtering the data to focus only on consistently tracked publishers over the entire 32-month period, we arrived at a balanced panel of 962 publishers (-88% change from the average of 8,334 publishers per month).

In contrast, the EU/US sample contains user location information, distinguishing between EU and US users. Still, it covers a shorter period, from April 2018 to December 2019 (21 months), and includes only a single pre-treatment period (April 2018). The unbalanced version of this sample includes an average of 7,264 publishers per month, with a maximum of 11,596 publishers in September 2018 and a minimum of 2,869 publishers in January 2019. We created a balanced panel of 717 publishers (-90% change from the average of 7,264 publishers per month) by selecting those consistently tracked throughout the 21 months in the EU/US sample. While this EU/US sample is not the focus of our main analysis, we use it in a robustness test, as outlined in the Web Appendix 9.2.

For the main analysis, we focused on the global sample because it offers a longer pre-treatment period, which is essential for robustly estimating the impact of the GDPR. We ensured consistency in our analysis by selecting the publishers that appeared in both the global and EU/US samples.

In both data sets, this step yielded 354 publishers (-63% change from the balanced global sample of 962 publishers).

To finalize the sample, we ensured that the assumption of parallel trends was fulfilled by examining the pre-treatment trends in the global sample. Specifically, we observed the pre-treatment trend of each publisher in the control group (in terms of the number of trackers). Then, we manually removed publishers from the control group with pre-treatment trends that were outliers compared to the treatment group. This process helped match the treatment and control groups' pre-treatment trends and ensured that the assumption of parallel trends was not violated. After removing these outliers, we arrived at a final balanced sample of 294 publishers (-17% change from the intersection of global and EU/US samples): 67 in the treatment group and 227 in the control group. Table 18 summarizes the steps to prepare the final sample for the analysis.

Table 18: Steps Taken to Prepare the Sample of 294 Publishers

| Step | Number of Publishers | Percent Change |
|---|---|---|
| Raw global sample (unbalanced; average number of publishers released monthly) | 8,334 | |
| Balanced global sample (May 2017 to December 2019) | 962 | -88.46% |
| Raw EU/US sample (unbalanced; average number of publishers released monthly) | 7,264 | |
| Balanced EU/US sample (April 2018 to December 2019) | 717 | -90.13% |
| Publishers present in both global and EU/US samples | 354 | -63.20% |
| Removing outliers in the control group of the global sample (ensuring parallel trends assumption) | 294 | -16.95% |

As shown in Table 18, we started with 962 publishers from the balanced global sample and 717 from the balanced EU/US sample. We filtered these publishers down to 354 (-63% change from the balanced global sample) by focusing on the ones in both samples. Finally, after removing outliers to ensure the parallel trends assumption held, we arrived at 294 publishers (-17% change from the intersection of global and EU/US samples). With a longer pre-treatment period, this sample forms the basis for our main analysis, while we use the EU/US sample as a robustness test in the Web Appendix 9.2.

### 9.9. Description of Publisher Industries and Types

Table 19: Description of Publisher Industries and Types

| Publisher Industry | | Publisher Type | Description of Publisher Type | Examples of Publishers |
|---|---|---|---|---|
| News Publishers | 1 | News & Portals | News providers and multipurpose portals, weather forecast publishers, TV channels, official publishers of cities, and sports associations. | bbc.com, accuweather.com, live.com |
| Non-News Publishers | 2 | E-Commerce | Shops whose websites allow purchasing items online without having a physical store. | amazon.com, nike.com, alibaba.com |
| | 3 | Recreation | Publishers where it is possible to book holidays or flights. | britishairways.com, booking.com, lufthansa.com |
| | 4 | Business | Publishers with a physical location and providing the option to purchase items online. Official company websites and publishers selling services fall within this category, too. | apple.com, paypal.com, airbnb.com |
| | 5 | Entertainment | Social networks and dating publishers, online games, video-sharing and streaming, and TV channels not focusing on news. | 9gag.com, facebook.com, spotify.com |
| | 6 | Reference | Search engines, wiki forums and communities, and online dictionaries. | google.com, wikipedia.org, dictionary.com |
| | 7 | Adult | Publishers that are generally thought not to be appropriate for children. | redtube.com, pornhub.com, sex.com |
| | 8 | Government | Official websites of political parties and movements. | nasa.gov, nih.gov, oevp.at |
| | 9 | Banking | Websites of banks. | americanexpress.com, deutsche-bank.de, commerzbank.de |

Notes: We adapt this table from Karaj et al. (2018a).

### 9.10. Robustness Test Regarding the Anticipation Assumption and the Influence of External Shocks in Early 2018

In this robustness test, we examine the possibility that publishers may have demonstrated a willingness to comply with the GDPR before its official enactment date (May 25[th], 2018) or with some delay. This analysis also addresses potential concerns about the influence of external shocks in early 2018. One significant event during this period was Facebook's Cambridge Analytica scandal, which broke in March 2018 (Cadwalladr and Graham-Harrison 2018). This scandal may have influenced publishers' use of trackers in ways unrelated to the GDPR. To account for such concerns, we exclude the months of March, April, May, and June 2018 from our main sample, thus testing the anticipation assumption and the impact of external shocks unrelated to the GDPR.

By removing these months, we aim to isolate the GDPR effect from any adjustments made by publishers in response to these earlier shocks or in anticipation of the GDPR. While including month-fixed effects helps control for broader time-related trends, excluding this specific time window offers an additional safeguard to ensure that early or late willingness to comply with the GDPR or external shocks unrelated to the regulation does not drive our main results. We run our baseline OLS regression specification from Equation (1) with this adjusted sample. Table 20 presents the results of this robustness test.

Table 20: Result of Difference-in-Differences Analysis without March, April, May and June 2018

| Dependent Variable: | Number of Trackers per Publisher and Month |
|---|---|
| Model: | (1) |
| Treatment x PostGDPR | -4.523* [-8.026; -1.020] |
| Publisher ID Fixed Effects | ✓ |
| Month ID Fixed Effects | ✓ |
| N Observations | 8,232 |
| R² | 0.741 |

Significance levels: * $p < 0.05$, ** $p < 0.01$, *** $p < 0.001$.
Two-way standard errors are clustered at the publisher and month levels; 95% confidence intervals are reported in brackets.
Notes: This table shows the difference-in-differences coefficient (Treatment x PostGDPR) from the OLS regression. We remove the months of March, April, May and June 2018 from our sample and assign treatment to each publisher according to the publisher's designation (EU or non-EU). Multiplying the number of publishers (N publishers = 294) and the number of months (T = 28 months) yields the number of observations (N observations = 8,232).

The DiD coefficient ($\beta$ = -4.523, $p < 0.05$, 95% CI [-8.026; -1.020]) is significantly negative. This result confirms that excluding March, April, May, and June 2018, which may reflect anticipation effects and the influence of the Cambridge Analytica scandal, does not alter our conclusion that the GDPR reduced the number of trackers by about 4 per publisher. Thus, early or late willingness to comply with the GDPR and external shocks unrelated to the GDPR do not bias our main findings.

### 9.11. Robustness Test Regarding the Skewness of the Dependent Variable

A potential concern with our DiD analysis is that the dependent variable, the number of trackers per publisher, is heavily right-skewed, which could bias its results. Figure 5 shows that most publishers typically use between 1 and 10 trackers, but a few use considerably more, with up to 111 trackers in a particular month. This skewed distribution of the dependent variable raises the concern that a few outliers might drive the DiD estimates, potentially biasing the main results.

To address this concern, we apply a log transformation to the dependent variable in this robustness test, which reduces the impact of extreme values and makes the distribution more symmetric. By doing so, we test whether our main findings hold when we account for the skewness in the dependent variable. Figure 11 shows the development of the log of the average monthly number of trackers in the treatment and control groups over time.

Figure 11: Development of the Log of Average Monthly Number of Trackers in the Treatment and Control Groups

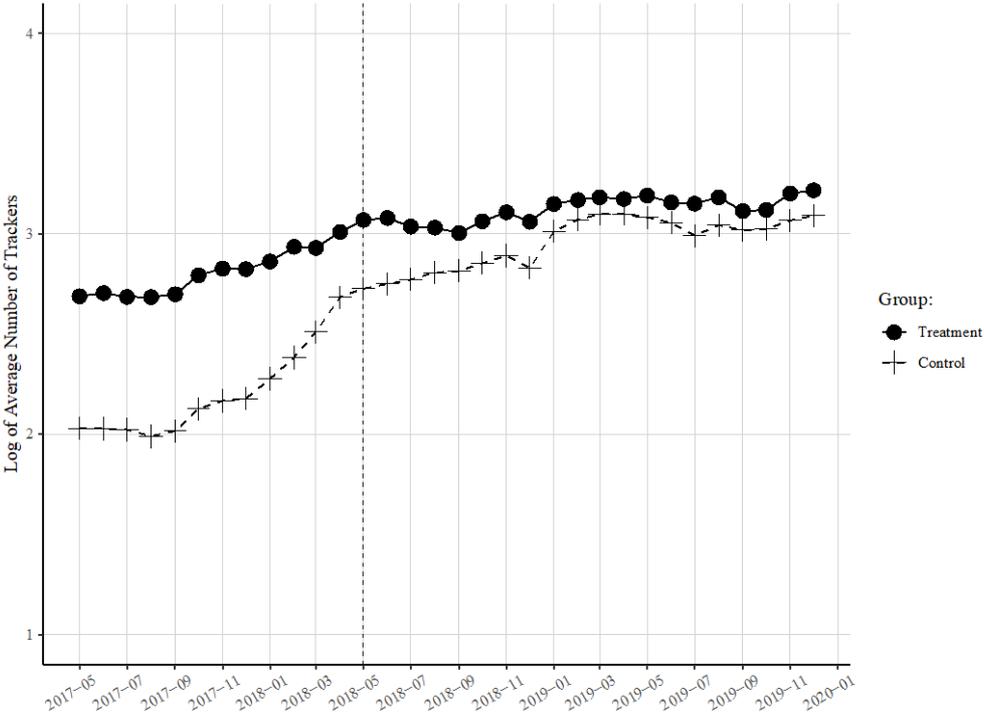

The parallel trends assumption holds with the log-transformed dependent variable, as the trends for both groups remain similar during the pre-treatment period. This result confirms that the log transformation does not violate the assumption of parallel trends. We run our baseline OLS regression specification from Equation (1) and present the results in Table 21.

Table 21: Result of Difference-in-Differences Analysis with Log Transformation of the Dependent Variable

| Dependent Variable: | Logarithm of the Number of Trackers per Publisher and Month |
|---|---|
| Model: | (1) |
| Treatment x PostGDPR | -0.490*** [-0.693; -0.287] |
| Publisher ID Fixed Effects | ✓ |
| Month ID Fixed Effects | ✓ |
| N Observations | 9,408 |
| $R^2$ | 0.749 |

Significance levels: * $p < 0.05$, ** $p < 0.01$, *** $p < 0.001$.
Two-way standard errors are clustered at the publisher and month levels; 95% confidence intervals are reported in brackets.
Notes: This table shows the difference-in-differences coefficient (Treatment x PostGDPR) from the OLS regression. We assign treatment to each publisher according to the publisher's designation (EU or non-EU). Multiplying the number of publishers (N publishers = 294) and the number of months (T = 32 months) yields the number of observations (N observations = 9,408).

The DiD coefficient is significantly negative ($\beta$ = -0.490, $p < 0.001$, 95% CI [-0.693; -0.287]). As we use the log of the dependent variable, we calculate the percentage change using the formula:

$$(e^{-0.490} - 1) \times 100 = (0.612 - 1) \times 100 = -38.74\%$$

This result indicates that the GDPR led to an approximate 38.74% reduction in the number of trackers per publisher after adjusting for the skewed distribution of the dependent variable.

In our main analysis, we estimate that the GDPR reduced the number of trackers by about 4 trackers per publisher, or 14.79% (see Table 6). The difference between these two results—the 38.74% reduction from the logged model and the 14.79% reduction from the main analysis—suggests that the log transformation–by dampening the impact of extreme values–reveals a stronger proportional reduction for the number of trackers.

This robustness test demonstrates that when accounting for skewness in the dependent variable, the effect of the GDPR on the number of trackers remains significantly negative. The log-transformed model results suggest an even larger relative decrease. This robustness test confirms that our main results are not sensitive to the skewness in the dependent variable. Even when using the log transformation, the assumption of parallel trends holds, and the effect of the GDPR on the number of trackers remains robust.

### 9.12. Detailed Report on the Estimated Fixed Effects

This section provides a detailed report on the fixed effects estimated from our baseline DiD regression specified in Equation (1). The fixed effects control for unobserved, time-invariant characteristics specific to publishers (publisher ID fixed effects) and common shocks affecting all publishers during specific months (month ID fixed effects).

Figure 12 presents the centered fixed effects coefficients for publisher ID and month ID. These effects were centered during estimation, making it easier to interpret the relative differences between publishers and months. The coefficients illustrate the variation in tracker usage across publishers and over time.

Figure 12: Estimated Fixed Effects Coefficients for Publisher ID and Month ID

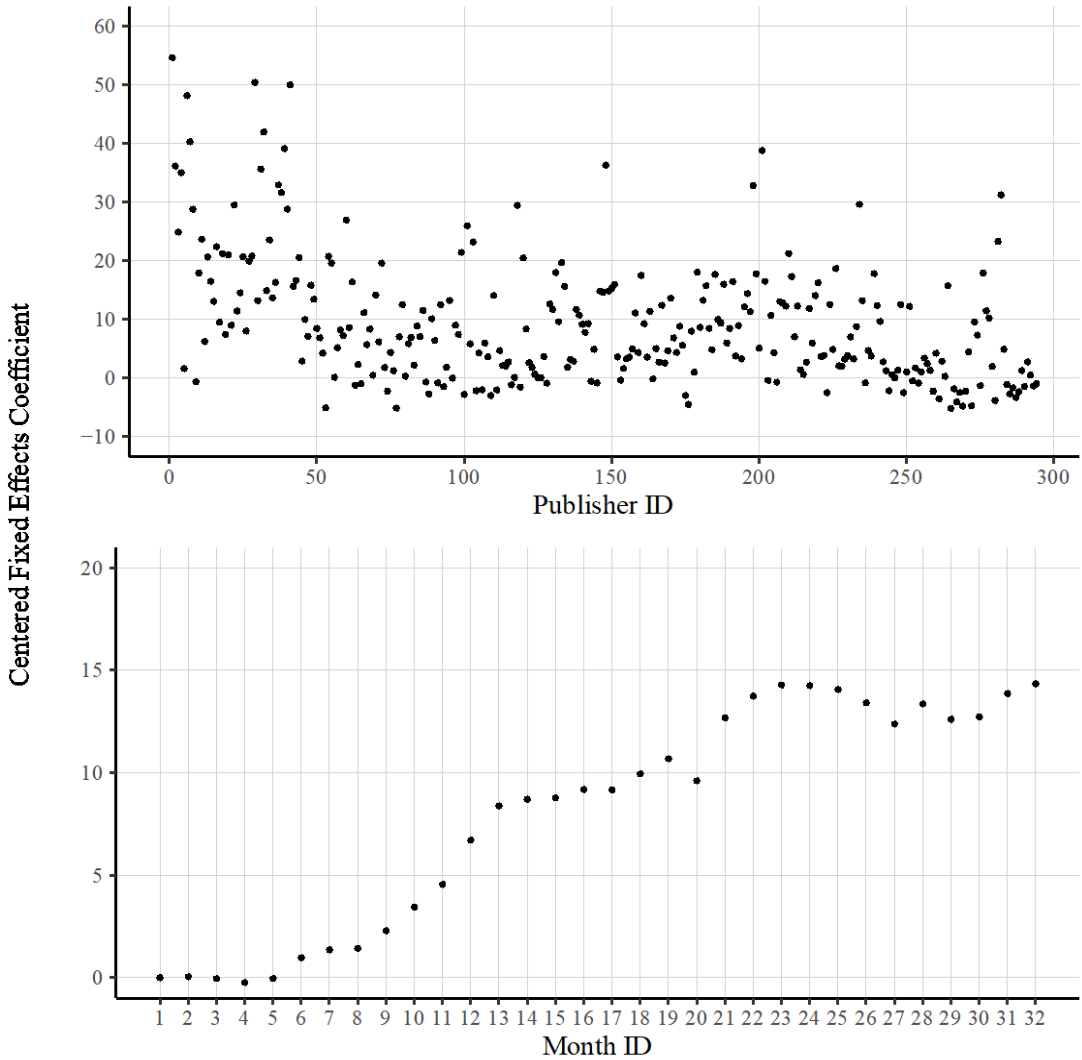

Figure 12 shows that most publishers have positive fixed effects, indicating higher-than-average tracker usage after accounting for other variables. The estimated month-fixed effects show an upward trend in publishers' tracker usage across months, suggesting a general, industry-wide increase in the adoption of trackers over time. While month-fixed effects control for time-specific shocks, such as seasonality or external events, these fixed effects may not fully capture this consistent upward trend.

To address this concern, we run an alternative specification of our baseline regression specification in Equation (6):

$$Y_{i,t} = \alpha + \gamma_i Treatment_i + \theta Time_t + \beta(Treatment_i \times PostGDPR_t) + \epsilon_{i,t} \qquad (6)$$

In Equation (6), we replace the month-fixed effects $\delta_t$ from Equation (1) with a linear time trend, denoted as $\theta Time_t$, where $Time_t$ represents the linear time trend for each month $t$. The time trend captures systematic changes in publisher tracker usage over time, independent of the GDPR. Other components of the regression remain the same as specified in Equation (1), with $\gamma_i$ accounting for publisher-fixed effects, and $\beta$ measuring the average treatment effect of the GDPR on the treated publishers (Table 22).

Table 22: Result of Difference-in-Differences Analysis for the Number of Trackers with a Linear Time Trend

| Dependent Variable: | Number of Trackers per Publisher and Month |
|---|---|
| Model: | (1) |
| PostGDPR | 4.404*** [2.474; 6.334] |
| Time | 0.356*** [0.260; 0.453] |
| Treatment x PostGDPR | -3.949* [-7.081; -0.817] |
| Publisher ID Fixed Effects | ✓ |
| N Observations | 9,408 |
| $R^2$ | 0.736 |

Significance levels: * $p < 0.05$, ** $p < 0.01$, *** $p < 0.001$.
Two-way standard errors are clustered at the publisher and month levels; 95% confidence intervals are reported in brackets.
Notes: This table shows the difference-in-differences coefficient (Treatment x PostGDPR) from the OLS regression. We assign treatment to each publisher according to the publisher's designation (EU or non-EU). Multiplying the number of publishers (N publishers = 294) and the number of months (T = 32 months) yields the number of observations (N observations = 9,408).

As seen in Table 22, including the linear time trend does not substantially alter our results. The DiD coefficient remains significantly negative ($\beta$ = -3.949, $p < 0.05$, 95% CI [-7.081; -0.817]), consistent with the results from our baseline specification. This result confirms that the GDPR's effect on treated publishers is robust, even after accounting for the general upward trend in tracker usage unrelated to the GDPR. The fact that the DiD coefficient is nearly identical in both models indicates that the general time trend is not masking the GDPR's effect. In other words, the baseline model with period fixed effects already captures the time variation well, and adding a linear trend does not change the main findings.

Lastly, while Figure 12 visually summarizes the estimated fixed effects, Table 23 presents the detailed numerical values of the estimated fixed effects from our baseline DiD regression specified in Equation (1). The mean-centered fixed effect for Publisher ID is 9.231, with a standard deviation of 10.774 across 294 publishers. Month ID's mean-centered fixed effect is 8.020, with a standard deviation of 5.423 across 32 months. These fixed effects capture systematic differences that are not attributable to the GDPR but are related to time-invariant characteristics of publishers and time-specific external factors.

Table 23: Detailed Estimated Fixed Effects Coefficients for Publisher ID and Month ID

| Type of FE | Publisher ID | Publisher ID Centered FE Coefficient | Month ID | Month ID Centered FE Coefficient |
|---|---|---|---|---|
| Publisher ID | 1 | 54.666 | | |
| Publisher ID | 2 | 36.136 | | |
| Publisher ID | 3 | 24.854 | | |
| Publisher ID | 4 | 35.011 | | |
| Publisher ID | 5 | 1.542 | | |
| Publisher ID | 6 | 48.198 | | |
| Publisher ID | 7 | 40.323 | | |
| Publisher ID | 8 | 28.761 | | |
| Publisher ID | 9 | -0.709 | | |
| Publisher ID | 10 | 17.855 | | |
| Publisher ID | 11 | 23.635 | | |
| Publisher ID | 12 | 6.167 | | |
| Publisher ID | 13 | 20.605 | | |
| Publisher ID | 14 | 16.448 | | |
| Publisher ID | 15 | 13.041 | | |
| Publisher ID | 16 | 22.355 | | |
| Publisher ID | 17 | 9.480 | | |
| Publisher ID | 18 | 21.167 | | |
| Publisher ID | 19 | 7.355 | | |
| Publisher ID | 20 | 20.980 | | |
| Publisher ID | 21 | 8.948 | | |
| Publisher ID | 22 | 29.480 | | |
| Publisher ID | 23 | 11.386 | | |
| Publisher ID | 24 | 14.480 | | |
| Publisher ID | 25 | 20.636 | | |
| Publisher ID | 26 | 7.948 | | |
| Publisher ID | 27 | 19.886 | | |
| Publisher ID | 28 | 20.792 | | |
| Publisher ID | 29 | 50.417 | | |
| Publisher ID | 30 | 13.136 | | |
| Publisher ID | 31 | 35.604 | | |
| Publisher ID | 32 | 41.979 | | |
| Publisher ID | 33 | 14.855 | | |
| Publisher ID | 34 | 23.511 | | |
| Publisher ID | 35 | 13.635 | | |
| Publisher ID | 36 | 16.260 | | |
| Publisher ID | 37 | 32.916 | | |
| Publisher ID | 38 | 31.605 | | |
| Publisher ID | 39 | 39.104 | | |
| Publisher ID | 40 | 28.791 | | |
| Publisher ID | 41 | 50.010 | | |
| Publisher ID | 42 | 15.573 | | |
| Publisher ID | 43 | 16.605 | | |
| Publisher ID | 44 | 20.511 | | |
| Publisher ID | 45 | 2.823 | | |
| Publisher ID | 46 | 9.917 | | |
| Publisher ID | 47 | 7.042 | | |
| Publisher ID | 48 | 15.761 | | |
| Publisher ID | 49 | 13.386 | | |

| Type of FE | Publisher ID | Publisher ID Centered FE Coefficient | Month ID | Month ID Centered FE Coefficient |
|---|---|---|---|---|
| Publisher ID | 50 | 8.355 | | |
| Publisher ID | 51 | 6.792 | | |
| Publisher ID | 52 | 4.167 | | |
| Publisher ID | 53 | -5.145 | | |
| Publisher ID | 54 | 20.698 | | |
| Publisher ID | 55 | 19.542 | | |
| Publisher ID | 56 | 0.042 | | |
| Publisher ID | 57 | 5.105 | | |
| Publisher ID | 58 | 8.135 | | |
| Publisher ID | 59 | 7.167 | | |
| Publisher ID | 60 | 26.916 | | |
| Publisher ID | 61 | 8.541 | | |
| Publisher ID | 62 | 16.355 | | |
| Publisher ID | 63 | -1.333 | | |
| Publisher ID | 64 | 2.229 | | |
| Publisher ID | 65 | -1.052 | | |
| Publisher ID | 66 | 11.104 | | |
| Publisher ID | 67 | 5.635 | | |
| Publisher ID | 68 | 8.291 | | |
| Publisher ID | 69 | 0.385 | | |
| Publisher ID | 70 | 14.105 | | |
| Publisher ID | 71 | 6.105 | | |
| Publisher ID | 72 | 19.542 | | |
| Publisher ID | 73 | 1.730 | | |
| Publisher ID | 74 | -2.333 | | |
| Publisher ID | 75 | 4.292 | | |
| Publisher ID | 76 | 1.167 | | |
| Publisher ID | 77 | -5.208 | | |
| Publisher ID | 78 | 6.979 | | |
| Publisher ID | 79 | 12.448 | | |
| Publisher ID | 80 | 0.230 | | |
| Publisher ID | 81 | 5.792 | | |
| Publisher ID | 82 | 6.886 | | |
| Publisher ID | 83 | 2.104 | | |
| Publisher ID | 84 | 8.823 | | |
| Publisher ID | 85 | 7.011 | | |
| Publisher ID | 86 | 11.479 | | |
| Publisher ID | 87 | -0.739 | | |
| Publisher ID | 88 | -2.802 | | |
| Publisher ID | 89 | 10.041 | | |
| Publisher ID | 90 | 6.323 | | |
| Publisher ID | 91 | -0.895 | | |
| Publisher ID | 92 | 12.448 | | |
| Publisher ID | 93 | -1.552 | | |
| Publisher ID | 94 | 1.761 | | |
| Publisher ID | 95 | 13.166 | | |
| Publisher ID | 96 | -0.052 | | |
| Publisher ID | 97 | 8.979 | | |
| Publisher ID | 98 | 7.355 | | |
| Publisher ID | 99 | 21.386 | | |

| Type of FE | Publisher ID | Publisher ID Centered FE Coefficient | Month ID | Month ID Centered FE Coefficient |
|---|---|---|---|---|
| Publisher ID | 100 | -2.864 | | |
| Publisher ID | 101 | 25.917 | | |
| Publisher ID | 102 | 5.760 | | |
| Publisher ID | 103 | 23.166 | | |
| Publisher ID | 104 | -2.239 | | |
| Publisher ID | 105 | 4.198 | | |
| Publisher ID | 106 | -2.083 | | |
| Publisher ID | 107 | 5.885 | | |
| Publisher ID | 108 | 3.542 | | |
| Publisher ID | 109 | -3.052 | | |
| Publisher ID | 110 | 14.042 | | |
| Publisher ID | 111 | -2.114 | | |
| Publisher ID | 112 | 4.605 | | |
| Publisher ID | 113 | 2.073 | | |
| Publisher ID | 114 | 1.948 | | |
| Publisher ID | 115 | 2.667 | | |
| Publisher ID | 116 | -1.208 | | |
| Publisher ID | 117 | 0.011 | | |
| Publisher ID | 118 | 29.417 | | |
| Publisher ID | 119 | -1.645 | | |
| Publisher ID | 120 | 20.480 | | |
| Publisher ID | 121 | 8.323 | | |
| Publisher ID | 122 | 2.573 | | |
| Publisher ID | 123 | 1.698 | | |
| Publisher ID | 124 | 0.573 | | |
| Publisher ID | 125 | -0.052 | | |
| Publisher ID | 126 | -0.052 | | |
| Publisher ID | 127 | 3.605 | | |
| Publisher ID | 128 | -0.958 | | |
| Publisher ID | 129 | 12.573 | | |
| Publisher ID | 130 | 11.636 | | |
| Publisher ID | 131 | 17.948 | | |
| Publisher ID | 132 | 9.542 | | |
| Publisher ID | 133 | 19.636 | | |
| Publisher ID | 134 | 15.573 | | |
| Publisher ID | 135 | 1.761 | | |
| Publisher ID | 136 | 3.105 | | |
| Publisher ID | 137 | 2.792 | | |
| Publisher ID | 138 | 11.636 | | |
| Publisher ID | 139 | 10.667 | | |
| Publisher ID | 140 | 9.136 | | |
| Publisher ID | 141 | 7.698 | | |
| Publisher ID | 142 | 9.229 | | |
| Publisher ID | 143 | -0.645 | | |
| Publisher ID | 144 | 4.823 | | |
| Publisher ID | 145 | -0.896 | | |
| Publisher ID | 146 | 14.761 | | |
| Publisher ID | 147 | 14.573 | | |
| Publisher ID | 148 | 36.291 | | |
| Publisher ID | 149 | 14.761 | | |

| Type of FE | Publisher ID | Publisher ID Centered FE Coefficient | Month ID | Month ID Centered FE Coefficient |
|---|---|---|---|---|
| Publisher ID | 150 | 15.230 | | |
| Publisher ID | 151 | 15.917 | | |
| Publisher ID | 152 | 3.542 | | |
| Publisher ID | 153 | -0.458 | | |
| Publisher ID | 154 | 1.573 | | |
| Publisher ID | 155 | 3.230 | | |
| Publisher ID | 156 | 3.511 | | |
| Publisher ID | 157 | 4.886 | | |
| Publisher ID | 158 | 11.041 | | |
| Publisher ID | 159 | 4.292 | | |
| Publisher ID | 160 | 17.448 | | |
| Publisher ID | 161 | 9.198 | | |
| Publisher ID | 162 | 3.511 | | |
| Publisher ID | 163 | 11.292 | | |
| Publisher ID | 164 | -0.208 | | |
| Publisher ID | 165 | 4.948 | | |
| Publisher ID | 166 | 2.604 | | |
| Publisher ID | 167 | 12.354 | | |
| Publisher ID | 168 | 2.480 | | |
| Publisher ID | 169 | 4.573 | | |
| Publisher ID | 170 | 13.573 | | |
| Publisher ID | 171 | 6.792 | | |
| Publisher ID | 172 | 4.292 | | |
| Publisher ID | 173 | 8.761 | | |
| Publisher ID | 174 | 5.511 | | |
| Publisher ID | 175 | -3.052 | | |
| Publisher ID | 176 | -4.583 | | |
| Publisher ID | 177 | 7.917 | | |
| Publisher ID | 178 | 0.916 | | |
| Publisher ID | 179 | 18.011 | | |
| Publisher ID | 180 | 8.573 | | |
| Publisher ID | 181 | 13.198 | | |
| Publisher ID | 182 | 15.698 | | |
| Publisher ID | 183 | 8.416 | | |
| Publisher ID | 184 | 4.761 | | |
| Publisher ID | 185 | 17.636 | | |
| Publisher ID | 186 | 9.948 | | |
| Publisher ID | 187 | 9.323 | | |
| Publisher ID | 188 | 15.917 | | |
| Publisher ID | 189 | 5.885 | | |
| Publisher ID | 190 | 8.355 | | |
| Publisher ID | 191 | 16.386 | | |
| Publisher ID | 192 | 3.698 | | |
| Publisher ID | 193 | 8.886 | | |
| Publisher ID | 194 | 3.198 | | |
| Publisher ID | 195 | 12.073 | | |
| Publisher ID | 196 | 14.355 | | |
| Publisher ID | 197 | 11.261 | | |
| Publisher ID | 198 | 32.791 | | |
| Publisher ID | 199 | 17.729 | | |

| Type of FE | Publisher ID | Publisher ID Centered FE Coefficient | Month ID | Month ID Centered FE Coefficient |
|---|---|---|---|---|
| Publisher ID | 200 | 5.042 | | |
| Publisher ID | 201 | 38.791 | | |
| Publisher ID | 202 | 16.448 | | |
| Publisher ID | 203 | -0.489 | | |
| Publisher ID | 204 | 10.635 | | |
| Publisher ID | 205 | 4.261 | | |
| Publisher ID | 206 | -0.770 | | |
| Publisher ID | 207 | 12.980 | | |
| Publisher ID | 208 | 12.761 | | |
| Publisher ID | 209 | 12.198 | | |
| Publisher ID | 210 | 21.166 | | |
| Publisher ID | 211 | 17.261 | | |
| Publisher ID | 212 | 6.980 | | |
| Publisher ID | 213 | 12.230 | | |
| Publisher ID | 214 | 1.292 | | |
| Publisher ID | 215 | 0.542 | | |
| Publisher ID | 216 | 2.636 | | |
| Publisher ID | 217 | 11.792 | | |
| Publisher ID | 218 | 5.886 | | |
| Publisher ID | 219 | 14.011 | | |
| Publisher ID | 220 | 16.198 | | |
| Publisher ID | 221 | 3.605 | | |
| Publisher ID | 222 | 3.791 | | |
| Publisher ID | 223 | -2.583 | | |
| Publisher ID | 224 | 12.480 | | |
| Publisher ID | 225 | 4.823 | | |
| Publisher ID | 226 | 18.636 | | |
| Publisher ID | 227 | 2.011 | | |
| Publisher ID | 228 | 1.917 | | |
| Publisher ID | 229 | 3.136 | | |
| Publisher ID | 230 | 3.730 | | |
| Publisher ID | 231 | 6.948 | | |
| Publisher ID | 232 | 3.198 | | |
| Publisher ID | 233 | 8.698 | | |
| Publisher ID | 234 | 29.605 | | |
| Publisher ID | 235 | 13.136 | | |
| Publisher ID | 236 | -0.864 | | |
| Publisher ID | 237 | 4.605 | | |
| Publisher ID | 238 | 3.667 | | |
| Publisher ID | 239 | 17.761 | | |
| Publisher ID | 240 | 12.323 | | |
| Publisher ID | 241 | 9.605 | | |
| Publisher ID | 242 | 2.698 | | |
| Publisher ID | 243 | 1.167 | | |
| Publisher ID | 244 | -2.239 | | |
| Publisher ID | 245 | 0.480 | | |
| Publisher ID | 246 | -0.052 | | |
| Publisher ID | 247 | 1.261 | | |
| Publisher ID | 248 | 12.480 | | |
| Publisher ID | 249 | -2.583 | | |

| Type of FE | Publisher ID | Publisher ID Centered FE Coefficient | Month ID | Month ID Centered FE Coefficient |
|---|---|---|---|---|
| Publisher ID | 250 | 0.948 | | |
| Publisher ID | 251 | 12.166 | | |
| Publisher ID | 252 | -0.552 | | |
| Publisher ID | 253 | 1.636 | | |
| Publisher ID | 254 | -0.927 | | |
| Publisher ID | 255 | 0.948 | | |
| Publisher ID | 256 | 3.323 | | |
| Publisher ID | 257 | 2.355 | | |
| Publisher ID | 258 | 1.230 | | |
| Publisher ID | 259 | -2.333 | | |
| Publisher ID | 260 | 4.167 | | |
| Publisher ID | 261 | -3.645 | | |
| Publisher ID | 262 | 2.791 | | |
| Publisher ID | 263 | 0.198 | | |
| Publisher ID | 264 | 15.698 | | |
| Publisher ID | 265 | -5.270 | | |
| Publisher ID | 266 | -1.927 | | |
| Publisher ID | 267 | -4.145 | | |
| Publisher ID | 268 | -2.552 | | |
| Publisher ID | 269 | -4.864 | | |
| Publisher ID | 270 | -2.333 | | |
| Publisher ID | 271 | 4.386 | | |
| Publisher ID | 272 | -4.833 | | |
| Publisher ID | 273 | 9.511 | | |
| Publisher ID | 274 | 7.260 | | |
| Publisher ID | 275 | -1.364 | | |
| Publisher ID | 276 | 17.854 | | |
| Publisher ID | 277 | 11.417 | | |
| Publisher ID | 278 | 10.167 | | |
| Publisher ID | 279 | 1.917 | | |
| Publisher ID | 280 | -3.927 | | |
| Publisher ID | 281 | 23.261 | | |
| Publisher ID | 282 | 31.198 | | |
| Publisher ID | 283 | 4.823 | | |
| Publisher ID | 284 | -1.177 | | |
| Publisher ID | 285 | -2.770 | | |
| Publisher ID | 286 | -1.739 | | |
| Publisher ID | 287 | -3.395 | | |
| Publisher ID | 288 | -2.427 | | |
| Publisher ID | 289 | 1.198 | | |
| Publisher ID | 290 | -1.520 | | |
| Publisher ID | 291 | 2.667 | | |
| Publisher ID | 292 | 0.448 | | |
| Publisher ID | 293 | -1.458 | | |
| Publisher ID | 294 | -1.052 | | |
| Month ID | | | 1 | 0.000 |
| Month ID | | | 2 | 0.044 |
| Month ID | | | 3 | -0.048 |
| Month ID | | | 4 | -0.238 |
| Month ID | | | 5 | -0.044 |

| Type of FE | Publisher ID | Publisher ID Centered FE Coefficient | Month ID | Month ID Centered FE Coefficient |
|---|---|---|---|---|
| Month ID | | | 6 | 0.973 |
| Month ID | | | 7 | 1.364 |
| Month ID | | | 8 | 1.429 |
| Month ID | | | 9 | 2.286 |
| Month ID | | | 10 | 3.442 |
| Month ID | | | 11 | 4.548 |
| Month ID | | | 12 | 6.707 |
| Month ID | | | 13 | 8.383 |
| Month ID | | | 14 | 8.706 |
| Month ID | | | 15 | 8.774 |
| Month ID | | | 16 | 9.189 |
| Month ID | | | 17 | 9.158 |
| Month ID | | | 18 | 9.948 |
| Month ID | | | 19 | 10.682 |
| Month ID | | | 20 | 9.611 |
| Month ID | | | 21 | 12.682 |
| Month ID | | | 22 | 13.747 |
| Month ID | | | 23 | 14.284 |
| Month ID | | | 24 | 14.254 |
| Month ID | | | 25 | 14.063 |
| Month ID | | | 26 | 13.420 |
| Month ID | | | 27 | 12.376 |
| Month ID | | | 28 | 13.366 |
| Month ID | | | 29 | 12.618 |
| Month ID | | | 30 | 12.723 |
| Month ID | | | 31 | 13.866 |
| Month ID | | | 32 | 14.342 |

Notes: We abbreviate "Fixed Effect" to "FE".
Summary of Centered Fixed Effects Coefficients:
- Publisher ID Centered FE Coefficients: Mean = 9.231, SD = 10.774, N = 294;
- Month ID Centered FE Coefficients: Mean = 8.020, SD = 5.423, N = 32.

## 9.13. Robustness Test Regarding the Consistency of EU Traffic Shares Across Public and Proprietary SimilarWeb Data Sets

In this section, we address the concern that using a single period of public SimilarWeb data set collected in September 2021 could result in the misclassification of publishers into treatment (EU) and control (non-EU) groups due to potential fluctuations in traffic shares over time. To ensure the robustness of our EU vs. non-EU publisher classifications, we validate the EU traffic shares from the public SimilarWeb data set against those from our proprietary SimilarWeb data set. The proprietary data set provides daily-level traffic shares for a subset of 200 out of 294 publishers (68%) in the balanced panel between January 2018 and December 2019.

For this comparison, we aggregated the post-GDPR periods from the proprietary data into a single post-GDPR period to directly compare with the EU traffic shares from the public SimilarWeb data collected in September 2021. We summarize the results of this comparison in Table 24, which shows the differences in EU traffic shares between the two data sets.

Table 24: Summary of EU Traffic Share Differences Between Public and Proprietary SimilarWeb Data Sets

| Statistic | Value (pp) |
|---|---|
| Average EU Traffic Share Difference | 11.08 pp |
| Max EU Traffic Share Difference | 60.39 pp |
| Min EU Traffic Share Difference | 0.00 pp |
| SD EU Traffic Share Difference | 10.37 pp |

Notes: This table summarizes the differences between the EU traffic shares in public and proprietary SimilarWeb data sets. The proprietary SimilarWeb data set spans multiple months pre- and post-GDPR (January 2018 - December 2019; T = 24 months). For this analysis, we aggregate the post-GDPR months in the proprietary SimilarWeb data set (T = 19 months) into a single post-GDPR period. The public SimilarWeb data set was collected in September 2021 (i.e., in the post-GDPR period; T = 1 month). We calculate the differences by comparing the aggregated post-GDPR EU traffic shares from proprietary SimilarWeb data set with the EU traffic shares in the public SimilarWeb data set for publishers that appear in both datasets (N publishers = 200). The average (mean) difference represents the average discrepancy in EU traffic shares between the two data sets across 200 publishers. The standard deviation indicates the variability of these differences. The total number of publishers (i.e., observations) included in the analysis is 200. All differences are expressed in percentage points (pp).

The average difference of 11.08 pp indicates that the traffic classifications as EU are broadly consistent across both data sources. The maximum difference of 60.39 pp reveals the largest discrepancy observed for any single publisher. The minimum difference of 0.00 pp shows that, for some publishers, the EU traffic share was identical across both data sets. The standard deviation of 10.37 pp suggests moderate variation in differences.

This analysis demonstrates that using a single period from the public SimilarWeb data set is mainly consistent with the results obtained from the proprietary data set across multiple periods. The differences observed, including the highest difference of 60.39 pp, may be partly due to the proprietary data's exclusion of major non-EU countries like Russia, which could potentially lower the non-EU traffic share calculations. However, despite this limitation, the overall differences in EU traffic shares remain relatively small, suggesting that our EU vs. non-EU classifications based on the public SimilarWeb data are robust.

We focus on EU traffic shares in this analysis because they directly relate to the study's primary goal of assessing the impact of the GDPR, an EU regulation, and ensuring the accuracy of our EU vs. non-EU publisher classifications. Additionally, the non-EU traffic shares in the proprietary data are less comprehensive, as they predominantly capture traffic from the United States, omitting significant non-EU countries like Russia. Including these non-EU countries in the proprietary data could introduce biases, making the EU traffic shares a more reliable metric for this comparison.

### 9.14. Robustness Test Regarding the Generalized Synthetic Control Method

One potential concern with our main difference-in-differences (DiD) analysis is the appropriateness of the two-way fixed effects estimator (TWFE) in capturing the average treatment effect on the treated (ATT). We use the Generalized Synthetic Control (GSC) method as a robustness test to address this concern (Xu 2017). GSC allows us to construct a counterfactual (i.e., a control group) by matching treated and control units more precisely and can serve as a robustness check to ensure that model misspecifications inherent in DiD analysis do not bias our main results.

In this robustness test, we use data from 354 publishers followed over 32 months (May 2017 to December 2019). This sample includes 67 publishers in the treatment group, consisting of EU-designated publishers subject to the GDPR, and 287 publishers in the control group, consisting of non-EU-designated publishers. Unlike in the DiD analysis, where we manually removed publishers from the control group with pre-treatment trends that were outliers compared to the overall treatment group's pre-treatment trend, we now allow the GSC method to select the optimal counterfactual using the pre-treatment data.

We do not use any covariates in constructing the counterfactual, as our data contains no time-varying covariates of the publishers. We construct the counterfactual solely based on the outcome variable (i.e., the number of trackers per publisher). This approach ensures that the observed trends between treatment and control groups align closely before the enactment of the GDPR, fulfilling (i.e., forcing) the assumption of parallel trends.

Using the expectation-maximization (EM) algorithm, the GSC method leverages pre-treatment information from the treatment group to construct the counterfactual, improving precision in estimates (Gobillon and Magnac 2016). We use cross-validation to select the optimal number of latent factors and choose five latent factors. The model includes two-way fixed effects (i.e., publisher- and month-fixed effects) to control for unobserved heterogeneity across publishers and months. In Figure 13, we compare the observed outcomes for the treatment group with the predicted counterfactual constructed using the GSC method (i.e., control group).

Figure 13: Development of the Average Monthly Number of Trackers in the Observed (Treatment) and Predicted Counterfactual (Control) Groups

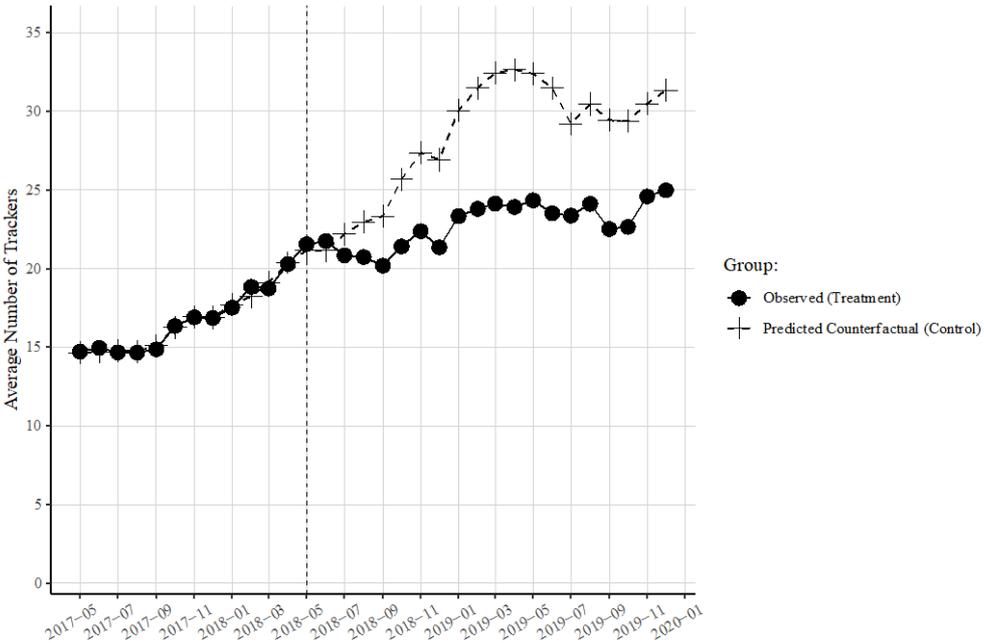

Notes: This figure shows the observed (treatment group) and predicted counterfactual (control group) number of trackers over time. We use the generalized synthetic control (GSC) method and the expectation–maximization (EM) algorithm to construct the predicted counterfactual.

Before the GDPR, the trends between the treatment and control groups closely align, satisfying the parallel trends assumption. After the GDPR, the trends between the two groups diverge. The number of trackers in the treatment group declines gradually after the GDPR. Shortly after that, the number of trackers in the treatment group increases and remains constant until the end of the observation period. In contrast, the predicted counterfactual (i.e., control group) shows a steady increase in trackers until April 2019. This difference between the observed and predicted counterfactual outcomes highlights the effect of the GDPR in reducing the number of trackers for the treatment group. Table 25 summarizes the estimated ATT coefficients across all months and by each month.

Table 25: Result of Generalized Synthetic Control Method Analysis for the Number of Trackers

| Dependent Variable: | Number of Trackers per Publisher and Month | |
|---|---|---|
| Model: | (1) | |
| | Estimate and 95% CI | |
| Average Treatment Effect on the Treated (ATT) Across All Months | -5.303*** | [-5.693; -4.914] |
| *ATT by Month:* | | |
| 2017-05 | 0.079 | [-0.386; 0.545] |
| 2017-06 | 0.272 | [-0.041; 0.586] |
| 2017-07 | -0.129 | [-0.496; 0.239] |
| 2017-08 | -0.075 | [-0.465; 0.315] |
| 2017-09 | -0.232 | [-0.657; 0.194] |
| 2017-10 | 0.069 | [-0.391; 0.529] |
| 2017-11 | -0.025 | [-0.403; 0.354] |
| 2017-12 | -0.048 | [-0.375; 0.280] |
| 2018-01 | -0.198 | [-0.657; 0.260] |
| 2018-02 | 0.619*** | [0.318; 0.919] |
| 2018-03 | -0.377 | [-0.806; 0.051] |
| 2018-04 | -0.086 | [-0.305; 0.132] |
| 2018-05 | 0.395 | [-0.755; 1.545] |
| 2018-06 | 0.585 | [-0.916; 2.087] |
| 2018-07 | -1.353 | [-2.945; 0.239] |
| 2018-08 | -2.220** | [-3.888; -0.552] |
| 2018-09 | -3.147*** | [-4.914; -1.380] |
| 2018-10 | -4.254*** | [-6.137; -2.370] |
| 2018-11 | -4.989*** | [-7.298; -2.680] |
| 2018-12 | -5.576*** | [-8.241; -2.911] |
| 2019-01 | -6.724*** | [-8.420; -5.027] |
| 2019-02 | -7.681*** | [-9.676; -5.686] |
| 2019-03 | -8.310*** | [-9.979; -6.642] |
| 2019-04 | -8.715*** | [-10.007; -7.424] |
| 2019-05 | -8.048*** | [-9.325; -6.771] |
| 2019-06 | -7.975*** | [-9.036; -6.913] |
| 2019-07 | -5.823*** | [-6.740; -4.906] |
| 2019-08 | -6.346*** | [-7.571; -5.120] |
| 2019-09 | -6.910*** | [-8.516; -5.303] |
| 2019-10 | -6.736*** | [-8.363; -5.108] |
| 2019-11 | -5.887*** | [-7.557; -4.217] |
| 2019-12 | -6.356*** | [-8.108; -4.605] |

Number of Observations: 11,328.
Number of Publishers in Treatment Group: 67; Number of Publishers in Control Group: 287.
95% confidence intervals are reported in brackets.
Significance levels: * $p < 0.05$, ** $p < 0.01$, *** $p < 0.001$.
Notes: This table presents results from the generalized synthetic control (GSC) method using the expectation-maximization (EM) algorithm. The EM algorithm leverages information from the treatment group during the pre-treatment period to improve the precision of estimates. The model includes two-way fixed effects (i.e., publisher and month-fixed effects). We use cross-validation to select the optimal number of latent factors and choose 5 latent factors. We perform parametric inference and use 1,000 bootstrap replications to estimate the confidence intervals. The model achieved a mean squared prediction error (MSPE) of 10.62, with an information criterion (IC) of 4.23 and a predictive criterion (PC) of 15.55. The residual variance was estimated at 8.4, and the EM algorithm converged after 37 iterations. We assign treatment to each publisher according to the publisher's designation (EU or non-EU). Multiplying the number of publishers (N publishers = 354) and the number of months (T = 32) yields the number of observations (N observations = 11,328).

The ATT coefficient (-5.303, $p < 0.05$, 95% CI [-5.682; -4.925]) is significantly negative in our GSC model presented in the column (1). This result is consistent with the findings from the DiD analysis (see Table 7). Lastly, Figure 14 plots the estimated ATT coefficients by each month – present in Table 25 – over time.

Figure 14: Development of the Monthly Average Treatment Effect (ATT) Coefficients for the Number of Trackers

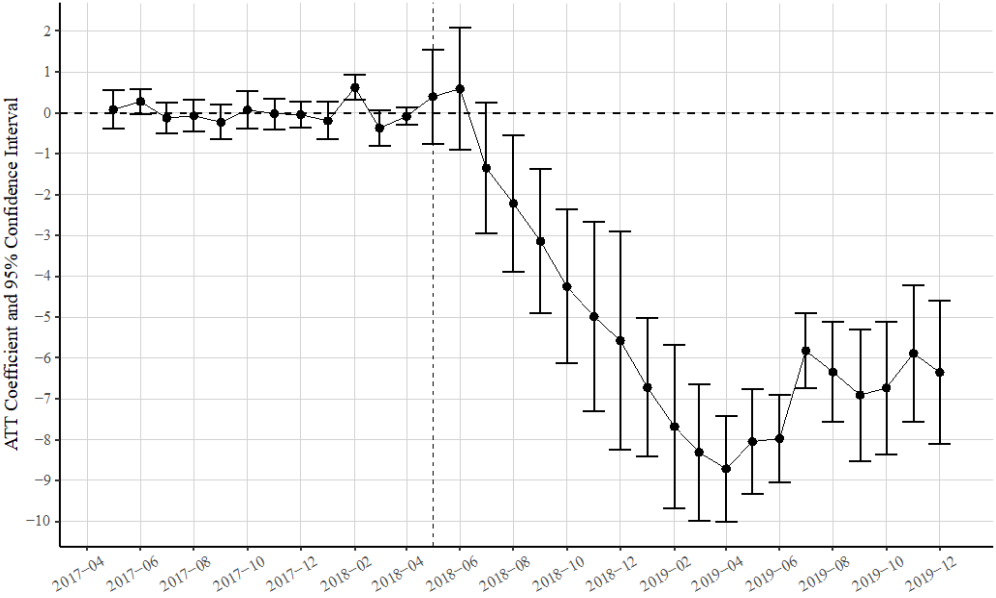

Notes: This figure shows the monthly average treatment effect (ATT) coefficients from the generalized synthetic control (GSC) method with the expectation–maximization (EM) algorithm. We assign treatment to each publisher according to the publisher's designation (EU or non–EU). The model leverages pre–treatment information from the treatment group to improve estimate precision and includes publisher and month fixed effects. Confidence intervals are based on parametric inference with 1,000 bootstrap replications.

These results demonstrate a significant negative effect of the GDPR on the number of trackers, with the ATT becoming progressively more negative over time. By the end of the observation period in December 2019, the ATT coefficient reaches -6.356 trackers (95% CI [-8.081; -4.632]), indicating a reduction in the number of trackers for the treatment group compared to the predicted counterfactual (i.e., control group).

The results from the GSC method's analysis confirm the findings from our main DiD analysis, showing a significant reduction in the number of trackers for publishers subject to the GDPR. The predicted counterfactual constructed using the GSC method provides additional confidence that potential model misspecifications in the DiD analysis do not bias our main result. The estimated ATT coefficients demonstrate the growing impact of the GDPR, with a gradual reduction in the number of trackers over the observation period. These results strengthen our conclusion that the GDPR significantly reduced the number of trackers for EU-designated publishers.

### 9.15.  Robustness Test Regarding the Unbalanced Panel

Our main analysis uses a balanced panel of publishers to ensure consistency over time and avoid panel attrition. WhoTracks.me released more publishers in their data sets over time, which could have led to an artificial increase in the number of trackers due to the inflow of new publishers with different tracker usage. Using a balanced panel, we ensure that each publisher is observed for the entire study period, preventing such distortions and ensuring that any changes in tracker usage are due to the GDPR's impact and not shifts in the sample composition.

In addition to avoiding panel attrition, the balanced panel simplifies the analysis by providing consistency across time, making it easier to interpret results. Every publisher is observed for the same period, which helps avoid issues related to missing data and ensures that each publisher contributes equally to the analysis. This approach allows us to focus on changes in the publisher's tracker usage across time without the added complexity of handling inconsistent data points.

However, we recognize the value of testing our results on a more representative sample of publishers. We use an unbalanced panel in this robustness test, allowing us to include more publishers in the sample even if WhoTracks.me did not observe them every month. Additionally, while the main analysis relied on a combination of top-level domain (TLD) and SimilarWeb traffic shares to classify publishers as EU or non-EU, collecting SimilarWeb traffic data for this expanded set of publishers is unfeasible for us. As such, we use TLD alone for treatment assignment in this robustness test, alongside an alternative treatment assignment based on server location. In the Web Appendix 9.1, we demonstrated that assigning treatment based on server location provides results consistent with our main findings, which reassures us of the validity of this alternative method to assign treatment.

This robustness test uses an unbalanced panel with 29,735 unique publishers and 256,595 observations over 32 months (compared to the 294 publishers and 9,408 observations in the balanced panel). We assign treatment using two approaches: (a) the publisher's TLD and (b) the publisher's server location. As we cannot collect the updated publisher website traffic data from SimilarWeb for this larger set of publishers, we rely solely on TLD to classify EU and non-EU publishers in this test. Figure 15 illustrates the development of the number of trackers for treatment and control groups under the two treatment assignment methods (TLD and server location).

Figure 15: Development of the Average Monthly Number of Trackers in the Treatment and Control Groups of the Unbalanced Panel

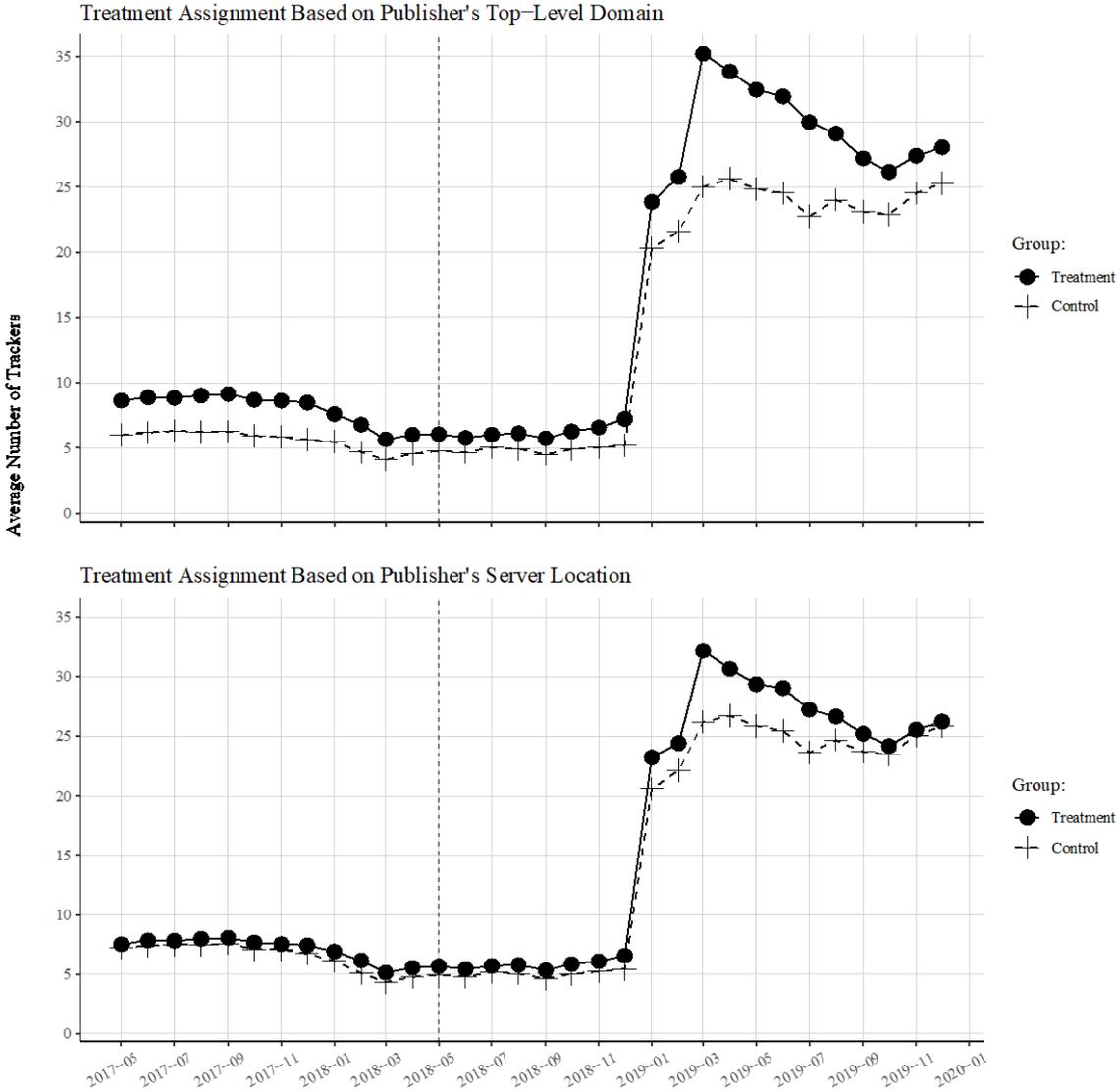

The trends suggest that the parallel trends assumption holds in the unbalanced panel. We observed a sharp increase in the number of trackers for both groups starting in January 2019, which coincides with a drop in the number of publishers included in the unbalanced panel.

Specifically, WhoTracks.me decreased the number of publishers from 14,194 in December 2018 to 3,015 in January 2019's data set release, representing a 78.76% decrease in the unbalanced panel. While this decrease may have affected the data collection process by WhoTracks.me, it is unclear whether this drop led to a sharp increase in the number of trackers for both groups starting in January 2019. We use the baseline OLS regression specified in Equation (1) to estimate the effect of the GDPR on the number of trackers in the unbalanced panel (Table 26).

Table 26: Result of Difference-in-Differences Analysis for the Number of Trackers in the Unbalanced Panel

| Dependent Variable: | Number of Trackers per Publisher and Month | |
|---|---|---|
| Model: | (1) | (2) |
| Treatment x PostGDPR | -1.081* [-2.122; -0.041] | -0.825* [-1.509; -0.141] |
| Publisher ID Fixed Effects | ✓ | ✓ |
| Month ID Fixed Effects | ✓ | ✓ |
| N Observations | 256,595 | 256,595 |
| $R^2$ | 0.781 | 0.781 |

Significance levels: * $p < 0.05$, ** $p < 0.01$, *** $p < 0.001$.
Two-way standard errors are clustered at the publisher and month levels; 95% confidence intervals are reported in brackets.
Notes: This table shows the difference-in-differences coefficient (Treatment x PostGDPR) from the OLS regressions. In model (1), we assign treatment to each publisher according to the publisher's top-level domain (EU or non-EU). In model (2), we assign treatment according to the publisher's server location (EU or non-EU). The panel includes 29,735 unique publishers (N publishers = 29,735) observed from May 2017 until December 2019 (T = 32 months), which would yield 951,520 observations in a balanced panel. However, because we use an unbalanced panel, the actual number of observations is lower (N observations = 256,595), reflecting the specific publisher-month combinations with available data.

The DiD coefficient ($\beta$ = -1.081, $p < 0.05$, 95% CI [-2.122; -0.041]) is significantly negative in our DiD model presented in column (1), where we assign the treatment to publishers based on TLD. When we assign the treatment to publishers based on server location, the DiD coefficient ($\beta$ = -0.825, $p < 0.05$, 95% CI [-1.509; -0.141]) is also significantly negative in column (2).

This robustness test demonstrates that the main findings from our analysis hold when using an unbalanced panel, which includes a larger number of publishers. The TLD and server location approach to the publisher's treatment assignment show consistent and significant reductions in the number of trackers after the GDPR.

### 9.16.  References in the Web Appendix